\documentstyle[aps,epsf]{revtex}  
\begin{document}  
\large  
\title {Relaxation to the Invariant Density for the Kicked Rotor}  
\author {Maxim Khodas  and Shmuel Fishman \\ \hskip.1in  
Department of Physics, Technion,   
Haifa 32000, Israel\\ Oded Agam\\ \hskip.1in The Racah  
Institute for Physics, The Hebrew  
University, Jerusalem 91904, Israel}  
\date{\today}  
\maketitle  
  
\begin{abstract}  
The relaxation rates to the invariant density in the chaotic phase space   
component of the kicked rotor (standard map) are calculated  
analytically for a large stochasticity parameter, $K$.   
These rates are the logarithms of the poles of the matrix elements of the   
resolvent, $ \hat{R}(z)=(z-\hat{U})^{-1} $, of the classical evolution   
operator $\hat{U} $. The resolvent poles are located  
inside the unit circle. For hyperbolic   
systems this is a rigorous result, but little is known about mixed   
systems such as the kicked rotor. In this work,  
the leading relaxation rates of the kicked rotor  
are calculated in presence of noise, to the leading order 
in $1/\sqrt{K}$. Then the limit of vanishing noise  is taken   
and the relaxation rates are found to be finite, corresponding to poles  
lying inside the unit circle. It is found that the slow relaxation  
rates, in essence, correspond to diffusion modes  
in the momentum direction. Faster relaxation modes 
intermix the motion in the momentum and the angle space. 
The slowest relaxation rate of distributions in the angle space 
is calculated analytically by studying the dynamics 
of inhomogeneities projected down to this space. 
The analytical results are verified by numerical simulations. 
\end{abstract}  
\ \ \ \ \ \ \ \ \ \ \ \ PACS number(s): 05.45.-a, 05.45.Ac  
\section{Introduction}  
For chaotic systems specific trajectories are extremely  
complicated and look random ~\cite{Ott}. Therefore it is natural   
to explore the statistical properties of such systems.   
For this purpose the evolution of probability densities of trajectories  
in phase space is studied ~\cite{Avez,Gaspardb,Dorfman}. For chaotic   
systems the probability  
densities, approach an equilibrium density that depends only on the system   
and not on the initial density. For hyperbolic systems (A systems), like  
the baker map, the relaxation   
is exponential. For such systems the existence of the relaxation  
rates   
was rigorously established  
and the relaxation rates are the Ruelle resonances ~\cite{Ruelle,CE,Gaspard}.  
To study these  
rates it is instructive to introduce the evolution operator of densities   
that is sometimes called the Frobenius-Perron (FP) operator.   
Relaxation to the equilibrium   
density is studied traditionally in statistical mechanics. In particular  
for particles performing a random walk in a finite box, relaxation to the   
equilibrium uniform density takes place and is governed by the   
rate related to the  
lowest nontrivial  
mode of diffusion equation. It is known that for the classical  
kicked rotor, described by the  standard map, diffusive spread in   
phase space takes place for a sufficiently large stochasticity parameter  
~\cite{1,2}.   
Therefore it is natural to study the Frobenius-Perron (FP) operator  
for the kicked rotor and to compare it to the diffusion operator, a  
comparison that enables to study some aspects of chaotic dynamics in the   
framework of statistical mechanics ~\cite{Balescu}. The kicked rotor model is   
a paradigm for the chaotic behavior of systems where one variable is  
unbounded in the phase space. For such classical systems diffusive spreading  
takes place.   
For their quantum counterparts  
it is suppressed by interference effects, leading to   
a mechanism that is similar to Anderson localization in disordered solids   
~\cite{7,10,Shmuel}.   
  
The kicked rotor is defined by the Hamiltonian (in appropriate units)  
  
\begin{equation}  
{\cal{H}}=\frac{1}{2}J^{2}+K cos\theta \sum_{n}\delta(t-n) ,   
\label{0}  
\end{equation}  
where $J$ is the angular momentum, $\theta$ is the conjugate angle  
$(0 \leq \theta < 2\pi)$ and $K$ is the stochasticity parameter.  
Since the angular momentum between the kicks is conserved, the equation   
of motion generated by (\ref{0}) reduces to a map, known as the standard map   
\begin{equation}  
\bar{\theta}=\theta+\bar{J}                  
\label{1}  
\end{equation}  
\begin{equation}  
\bar{J}=J-K sin\theta ,                  
\label{2}  
\end{equation}  
where $\theta$ and $J$ are the angle and the angular momentum before the kick,  
while $\bar{\theta}$ and $\bar{J}$ are these quantities just before the next   
kick. For $K > K_{c} \approx 0.9716$ diffusion in phase space   
is found, and for   
large $K$ the diffusion coefficient is given by an expansion in  
$\frac{1}{\sqrt{K}}$ as:  ~\cite{RW}  
\begin{equation}  
D(K)=\frac{K^{2}}{4} \left(1-2J_{2}(K)+... \right).  
\label{d1}  
\end{equation}  
To be precise it was shown that after a large number of kicks, n, the   
variance of the momentum behaves as:  
\begin{equation}  
\langle (J- \langle J \rangle)^{2} \rangle \sim  2Dn,  
\label{d2}  
\end{equation}  
where $ \langle \cdots \rangle$ denotes an averaging over the  
angle initial distribution, and  
 $D$  is given by  (\ref{d1}).  
  
It is assumed that the system evolves in presence of finite noise and the limit of  
the vanishing noise is taken in the end of the calculation. The noise is required  
here in order to get well defined results. It leads to escape from the accelerator  
modes and other stable islands. Accelerator modes, where the angular momentum  
$J$ grows linearly with time, are found for values of $K$ and the initial values  
($\theta_0$,$J_0$) of the angle and the angular momentum, that satisfy $Ksin  
\theta_0=2\pi l_0$ and $J=2\pi l$ where $l$ and $l_0$ are integers. In such a  
situation at each step, $J$ grows by $2\pi l_0$, as is obvious from  
(\ref{1}) and (\ref{2}), namely its growth is linear in time. For some values of $K$ the  
point ($\theta_0$,$J_0$) is stable and also for initial conditions in its vicinity   
the momentum grows linearly.  This differs from diffusion, that takes place  
in the chaotic component of phase space.  
For trajectories in the chaotic component of phase space noise avoids  
long time sticking in the vicinity of islands of stability ~\cite{Zaslav}.  
In numerical calculations   
without noise  
diffusion is found for $K>K_{c}$ for  
trajectories in the extended chaotic component for large values of  
$K$, but also some exceptions were reported ~\cite{Zaslav}.  
The diffusion coefficient   
(\ref{d1}) was calculated in   
presence of finite noise (in the long time limit) and the limit of the   
vanishing noise can be taken in the end of the calculation~\cite{RW}.   
It describes the   
typical spreading of trajectories in   
the chaotic component. Since the kicked rotor is a mixed system, as is   
the case  
for most physical examples, the rigorous mathematical theory for relaxation  
~\cite{Ruelle,Gaspardb,Avez} does not apply and one has to resort to   
heuristic methods.  
  
In the present paper \cite{PRL} the Frobenius-Perron operator will be calculated  
for  
the kicked rotor on the torus:   
\begin{equation}  
\ \ \ (0 \leq J < 2\pi s)\\  
\label{d3}  
\end{equation}  
\[  
(0 \leq \theta < 2\pi),  
\]  
where $s$ is integer.   
This is reasonable since the map (\ref{1}-\ref{2}) is $2\pi$ periodic both  
in $J$ and in  $\theta$. The operator is defined in the space spanned by the   
Fourier basis  
\begin{equation}  
\phi_{km}=(J\theta|km)=  
\frac{1}{\sqrt{2\pi}}\frac{1}{\sqrt{2\pi s}}  
\exp (im\theta)\exp \left( i\frac{kJ}{s} \right).  
\label{3}  
\end{equation}  
Note that the functions $\phi_{k0}$ form the basis of eigenstates of the   
diffusion operator in the angular momentum $J$.   
The FP operator for an area preserving invertible map,  
\[  
\bar{\bf x}=M({\bf x})  
\]  
is   
\begin{equation}  
\hat{U}\rho(\theta,J)=\rho(M^{-1}(\theta,J)).  
\label{4}  
\end{equation}  
It was studied rigorously for the hyperbolic systems and many of its  
properties are known ~\cite{Avez,Gaspardb,Ruelle,HS,Oded}. It is a  
unitary operator in ${\cal L}^{2}$,  
the Hilbert space of square integrable functions. Therefore its resolvent  
\begin{equation}  
\hat{R}(z)=\frac{1}{z-\hat{U}}=  
\frac{1}{z}\sum_{n=0}^{\infty}\hat{U}^{n}z^{-n}   
\label{5}  
\end{equation}  
is singular on the unit circle in the complex $z$ plane.  
The matrix elements of $\hat{R}$ are discontinuous there and one finds a jump  
between two Riemann sheets.   
This results from the fact that the spectrum is continuous and infinitely   
degenerate ~\cite{Avez}.  
The sum (\ref{5}) is convergent for $|z|>1$,  
therefore it identifies the physical sheet, as the one connected with the   
region  $|z|>1$. (This is analogous to the sign of the small imaginary   
increment  
in the energy that is used in the definition of the Green function). The  
Ruelle resonances are the poles of the matrix elements of the resolvent, on  
the Riemann sheet, extrapolated from $|z|>1$ ~\cite{HS}. These describe   
the decay of {\em smooth} probability distribution functions to the   
invariant density in a coarse grained form ~\cite{Gaspardb,Ruelle}.   
Even a smooth initial distribution will develop complicated patterns   
as a result   
of the evolution of a chaotic map. The Ruelle resonances describe   
the decay of its   
{\em coarse grained} form to the invariant density. In spite of the solid   
mathematical  
theory there are very few examples where 
the Ruelle resonances were calculated  
 for specific systems. They were calculated analytically for the baker map  
where the basis of Legendre polynomials was used 
~\cite{HS} and various of its variants ~\cite{Gaspardb}. 
The Ruelle resonances were calculated also for the ``cat'' map and some of   
its variants ~\cite{Oded}. 
Blum and Agam applied a variational method for the  
calculation of the leading Ruelle resonances of the ``perturbed cat'' map, 
and the results were verified numerically\cite{blum}.  
In addition they calculated the leading  
resonances of the standard map with $s=1$ for various values
of the stochasticity  
parameter $K$. The leading Ruelle resonances  for the kicked top
were calculated by    
Weber, Haake and \v{S}eba ~\cite{HW} with the help of a  
combination of a cycle expansion and numerical calculations.  
The resonances mentioned above are not related to the spectrum of the
Liouville operator that is
confined to the unit circle because of its unitarity.

In the present work,  the FP operator is calculated for the kicked rotor. 
Here the classical evolution operator, for one time step,  
can be written in the form  
\begin{equation}  
\hat{U}_{KR}=  
\delta(\bar{\theta}-\theta-\bar{J})\delta(\bar{J}-J+K sin\theta),  
\label{7}  
\end{equation}  
and its operation on a phase space density $\rho$ is  
\begin{equation}  
\label{6}  
\hat{U}\rho(\theta,J)=\rho(\theta-J,J+K sin(\theta-J)).  
\end{equation}  
To make the calculation well defined noise is added to the system. It is   
shown that the noise acts  
effectively as coarse graining and the resulting FP operator is not unitary   
(see also ~\cite{Shmuelinbook}). For large stochasticity  
parameter $K$,  we show that 
the slowest   
relaxation modes in the limit of infinitesimal noise are the   
modes of the diffusion operator in the angular momentum space.
Also calculated is the slowest rate of relaxation in the angle space.
The approximate analytical results are tested numerically. 
 
It is understood that the noise is kept finite when the limits of large   
$K$ and $s$ are taken   
and then the limit of zero noise is taken.   
The natural question is whether it is possible 
that this description, which was   
established only for hyperbolic systems, holds also for mixed systems.   
Clearly, for mixed systems it can be only    
approximate. It holds for large values of the stochasticity parameter   
$K$ since then most of the phase space is covered by the chaotic   
component. For smaller values of $K$ the weight of the regular regions   
increases. In such a situation, in the limit of increasing resolution the   
resonances related to the regular component   
are expected to move   
to the unit circle in the complex $z$ plane,   
corresponding to the quasi-periodic motion, while the resonances associated   
with the chaotic component stay inside the unit circle. This was found by   
Weber, Haake and \v{S}eba ~\cite{HW} for the kicked top that is a mixed system.  
  
How is the FP operator related to the quantum mechanical evolution operator?   
It was shown numerically for the baker map  
that if both operators are calculated with finite resolution they exhibit the   
same Ruelle  
resonances ~\cite{Shmuelinbook}. In this calculation it was assumed that the   
phase space   
coarse graining tends to zero in the semi-classical limit   
$\hbar \rightarrow 0$.   
It was shown by Zirnbauer ~\cite{Zir} that some noise is required   
for a meaningful  
definition of the field theories introduced to study level statistics for   
chaotic systems ~\cite{AAAS}.  
This noise affects only quantum properties, therefore the resulting   
ensemble has the same classical FP operator.  
The localization length of the kicked rotor 
calculated from this field theory ~\cite{Atland} is related 
to the classical FP operator. This operator is analyzed in the present work, clarifying some issues of that work.
The results hold only for typical   
quantum systems, since the noise introduced in the present work as well as   
the noise required for the stabilization of the field theory ~\cite{Zir}   
washes out  the sensitive quantum details such as the number  
theoretical properties of   
the effective Planck constant ~\cite{Casati}.  
  
The Frobenius-Perron operator in the basis (\ref{3}) in the presence of noise   
is defined and calculated in Sec. 2, its  
Ruelle resonances are obtained within some approximations in Sec. 3   
and their regime of validity is tested numerically in Sec. 4. The results are   
summarized and discussed in Sec. 5.

\section{The evolution operator of phase space distributions}  
  
In this section the evolution operator of phase space densities  
of the kicked rotor in presence of some type of noise is  
derived.   
The noise is added to the free motion part (\ref{1}). In the absence of noise  
the phase space evolution   
of a distribution $f$  
is given by Liouville equation  
\begin{equation}  
\frac{df}{dt}=\frac{\partial f}{\partial t}+  
\dot{\theta}\frac{\partial f}{\partial \theta}+\dot{J}\frac{\partial f}{\partial J}=0.   
\label{s1}  
\end{equation}  
If noise that conserves J, and leads to diffusion in $\theta$,  
is added to the free motion, equation (\ref{s1}) should be  
replaced by   
\begin{equation}  
\frac{\partial f}{\partial t}+  
J\frac{\partial f}{\partial \theta}-  
\frac{\sigma^{2}}{2}\frac{\partial^{2} f}{\partial \theta^{2}}=0,  
\label{s2}  
\end{equation}  
where $J=\dot{\theta}$ was used.  
It can be written as:   
\begin{equation}  
\frac{\partial f}{\partial t}=\hat{A}f,  
\label{s3}  
\end{equation}  
where the operator $\hat{A}$ is:  
\begin{equation}  
\hat{A}=-J\frac{\partial }{\partial \theta}+  
\frac{\sigma^{2}}{2}\frac{\partial^{2} }{\partial \theta^{2}}.  
\label{s4}  
\end{equation}  
The complete set of its eigenfunctions is given by   
$\varphi_{m}=\frac{1}{\sqrt{2\pi}} \exp(im\theta) $, where $m$ is integer.    
The operator we need is \ $ \hat{U}_{noise}=e^{\hat{A}}$, and explicitly  
\begin{equation}  
(J'\theta'|\hat{U}_{noise}|J\theta)=  
\sum_{m}^{} \varphi_{m}(\theta')\varphi_{m}^{*}(\theta)  
\exp(\alpha_{m}) \delta(J-J') ,    
\label{s5}  
\end{equation}  
where  the $\alpha_{m}$ are the eigenvalues of the operator $\hat{A}$,   
namely $\hat{A}|\varphi_{m})=\alpha_{m}|\varphi_{m})$.   
Obviously   
\begin{equation}  
\alpha_{m}=-imJ-\frac{\sigma^{2}}{2}m^{2}  
\label{s6}  
\end{equation}  
leading to  
\begin{equation}  
(J'\theta'|\hat{U}_{noise}|J\theta)=  
\sum_{m}^{} \frac{1}{2\pi}\exp \left(im(\theta'-\theta-J)-  
\frac{\sigma^{2}}{2}m^{2}\right) \delta(J-J').  
\label{s7}  
\end{equation}    
The $\delta$-function in momentum reflects the fact that the noise  
does not affect the momentum.   
The matrix elements   
$ (k_{2}m_{2}|\hat{U}|k_{1}m_{1}) $   
of evolution operator $\hat{U}$ in the Fourier basis (\ref{3})  
will be calculated in two steps, first the contribution of the kick, and then  
the one of the free motion with noise will be calculated.   
According to (\ref{2}) and (\ref{6}),  
the kick transforms the state   
$(\theta,J|k_{1}m_{1})=\frac{1}{\sqrt{2\pi}}\frac{1}{\sqrt{2\pi s}}  
\exp (im_{1}\theta)\exp \left(i\frac{k_{1}J}{s}\right)$   
to the state   
\begin{equation}  
\frac{1}{\sqrt{2\pi}}\frac{1}{\sqrt{2\pi s}}  
\exp (im_{1}\theta)\exp \left(i\frac{k_{1}}{s}(J+K sin \theta)\right)   
\equiv (\theta,J|\hat{U}_{K}|k_{1}m_{1}).  
\label{s8}  
\end{equation}    
Adding the effect of noise yields the matrix element in the  
mixed representation  
\begin{equation}  
(J\theta|\hat{U}|k_{1}m_{1})=   
\int_{0}^{2\pi}d\theta'\int_{0}^{2\pi s}dJ'  
(J\theta|\hat{U}_{noise}|J'\theta')(J'\theta'|\hat{U}_{K}|k_{1}m_{1}).   
\label{s9}  
\end{equation}  
Its transformation to the basis (\ref{3}) is calculated in App. A   
and the result is  
\begin{equation}  
(k_{2}m_{2}|\hat{U}|k_{1}m_{1})=  
J_{m_{2}-m_{1}}\left(\frac{k_{1}K}{s}\right)  
\exp \left(-\frac{\sigma^{2}}{2}m_{2}^{2}\right)  
\delta_{k_{2}-k_{1},m_{2}s}.    
\label{start}  
\end{equation}  
For $\sigma=0$, using ({\ref{start}) one can verify by a straightforward 
summation that  
$\hat{U}^{\dagger}\hat{U}=I$, therefore the operator is unitary as required.

Some of the eigenfunctions of $\hat{U}$ in the limit $\sigma=0$ are easily  
found. It is convenient to use the representation (\ref{7}) of $\hat{U}$.  
We guess an eigenfunction of the form  
\begin{equation}  
F(\theta,J)=\delta(\theta-\theta_{0})\sum_{l}^{}\exp(iqJ/s)\delta(J-2\pi l)  
\label{s13}  
\end{equation}  
with q integer satisfying $1\leq q \leq s$. These are functions   
localized on accelerator modes representing linear growth of the angular momentum   
with time.  
To check that these are indeed eigenfunctions we note that   
\begin{equation}  
\hat{U}_{KR}F(\theta,J)=  
\delta(\theta -J -\theta_{0})  
\sum_{l}^{}\delta(J+K sin\theta_{0}-2\pi l)  
\exp \left(i\frac{q}{s}(J+K sin\theta_{0})\right)  
\label{s14}  
\end{equation}  
taking $\theta_{0}$ so that   
\[  
K sin\theta_{0}=2\pi l_{0},  
\]  
where $l_{0}$ is integer yields  
\begin{equation}  
\hat{U}_{KR}F(\theta,J)=e^{i2\pi ql_{0}/s} F(\theta , J),  
\label{s15}  
\end{equation}  
since the RHS of (\ref{s14}) does not vanish only for $J=2\pi l $.  
The eigenvalues $e^{i2\pi ql_{0}/s}$ lie on the unit  
circle and become dense as $s \rightarrow \infty$.  
There are more eigenfunctions of this form located on other  
periodic orbits ~\cite{Berry}.

\section{Identification of the Ruelle resonances}  
The purpose of this section is to calculate the    
Ruelle resonances for the kicked rotor with the help of   
Frobineus-Perron operator (\ref{start}).   
The calculation will be done for finite noise $\sigma$   
and then the limit  $\sigma \rightarrow 0$ will be taken.  
The Ruelle resonances are the poles of matrix elements of  
the resolvent operator $\hat{R}$ of (\ref{5}),  
\begin{equation}  
R_{12}=(k_{1}m_{1}|\hat{R}(z)|k_{2}m_{2})=  
\left(k_{1}m_{1}\left|\frac{1}{z-\hat{U}}\right|k_{2}m_{2}\right),   
\label{u1}  
\end{equation}  
when analytically continued from outside of the unit circle   
in the complex plane.  
It is useful to introduce the operator  
\begin{equation}  
\hat{R}'(z)=\frac{1}{1-z\hat{U}},  
\label{u2}  
\end{equation}  
that is related to the resolvent by  
\begin{equation}  
\frac{1}{z}\hat{R}'\left(\frac{1}{z}\right)=\hat{R}(z)  
\label{u3}  
\end{equation}  
and  
\begin{equation}  
\frac{1}{z}\hat{R}\left(\frac{1}{z}\right)=\hat{R}'(z).  
\label{u4}  
\end{equation}  
The matrix elements of   
$\hat{R}$ and $\hat{R}'$  
satisfy similar relations.  
Continuing the matrix elements of $\hat{R}(z)$ from outside to inside of unit   
circle is equivalent to continuing the matrix elements of $\hat{R}'(z)$ from   
inside to outside of unit circle. The last continuation is easier to study   
since   
the expansion  
\begin{equation}  
\hat{R}'(z)=\frac{1}{1-z\hat{U}}=\sum_{n=0}^{\infty}z^{n}\hat{U}^{n}   
\label{u6}  
\end{equation}  
is convergent inside the unit circle, because  
\begin{equation}  
||z\hat{U}||\leq 1.  
\label{u7}  
\end{equation}  
The resulting matrix elements are  
\begin{equation}  
R^{'}_{12}=(k_{1}m_{1}|\hat{R}'(z)|k_{2}m_{2})=\sum_{n=0}^{\infty}a_{n}z^{n},  
\label{u8}  
\end{equation}  
where  
\begin{equation}  
a_{n}=(k_{1}m_{1}|\hat{U}^{n}|k_{2}m_{2}).  
\label{u9}  
\end{equation}  
Through (\ref{u3}) and (\ref{u4}) this expansion is related to matrix   
elements outside of the unit circle.   
If $z_{c}$ is a singularity of $R_{12}$ then $1/z_{c}$ is a singular  
point of $R^{'}_{12}$. Consequently  
the first singularity of the analytic continuation of  $R^{'}_{12}(z)$ from   
inside to outside the unit circle gives the   
first singularity one encounters when analytically continuing  $R_{12}(z)$ from   
outside to inside the unit circle, i.e. it is just the leading  
nontrivial resonance.   
This is the most interesting resonance determining the relaxation to the   
invariant density. The first singularity in the extrapolation of the  
matrix elements of $\hat{R}'$ from inside to outside the unit circle   
is determined   
from the fact that it is the radius of convergence of  
this series.   
Moreover according to the Cauchy-Hadamard theorem (see ~\cite{CA})   
the inverse of the radius of convergence is given by   
\begin{equation}  
r^{-1}=\lim_{n \rightarrow \infty}\sup \sqrt[n]{|a_{n}|}.  
\label{u10}  
\end{equation}  
If $a_{n} \sim c/r^{n}$ we may say that the  
radius of convergence is the asymptotic value of  $a_{n-1}/a_{n}$. This is    
the basis for the ratio method for determining the radius of convergence.  
The resonance that is closest to the unit circle can be identified   
from the radius of convergence.  
  
We turn now to calculate the coefficients $a_{n}$.  
First the matrix elements   
$(k0|\hat{R}'(z)|k0)$ will be calculated. For these the expansion   
coefficients are:   
\begin{equation}  
a_{n}=(k0|\hat{U}^{n}|k0).  
\label{u11}  
\end{equation}  
Introducing the resolution of the identity,  
\begin{equation}  
a_{n}=\sum_{k_{1},m_{1}}\sum_{k_{2},m_{2}}...  
\sum_{k_{n-1},m_{n-1}}  
(k0|\hat{U}|k_{1}m_{1})(k_{1}m_{1}|\hat{U}|k_{2}m_{2})...  
(k_{n-1}m_{n-1}|\hat{U}|k0),  
\label{u12}  
\end{equation}  
and substitution of (\ref{start}) leads to:  
\begin{equation}  
a_{n}=\sum_{k_{1},m_{1}}\sum_{k_{2},m_{2}}...  
\sum_{k_{n-1},m_{n-1}}  
J_{0-m_{1}}\left(\frac{k_{1}K}{s}\right)\delta_{k-k_{1},0}  
J_{m_{1}-m_{2}}\left(\frac{k_{2}K}{s} \right)  
\exp \left(-\frac{\sigma^{2}}{2}m_{1}^{2}\right)\delta_{k_{1}-k_{2},m_{1}s}  
\label{u13}  
\end{equation}  
\[  
J_{m_{2}-m_{3}}\left(\frac{k_{3}K}{s} \right)  
\exp \left(-\frac{\sigma^{2}}{2}m_{2}^{2}\right)\delta_{k_{2}-k_{3},m_{2}s}  
...J_{m_{n-1}-0}\left(\frac{kK}{s}\right)  
\exp \left(-\frac{\sigma^{2}}{2}m_{n-1}^{2}\right)\delta_{k_{n-1}-k,m_{n-1}s}.  
\]  
Summation over the $k_{i}$ yields,  
\begin{equation}  
a_{n}=\sum_{m_{1}}\sum_{m_{2}}...  
\sum_{m_{n-1}}  
J_{0-m_{1}}\left(\frac{kK}{s}\right)  
\label{u14}  
\end{equation}   
\[  
J_{m_{1}-m_{2}}\left(\frac{kK}{s}-m_{1}K \right)  
\exp \left(-\frac{\sigma^{2}}{2}m_{1}^{2}\right)  
J_{m_{2}-m_{3}}\left(\frac{kK}{s}-(m_{1}+m_{2})K \right)  
\exp\left(-\frac{\sigma^{2}}{2}m_{2}^{2}\right)   
\]  
\[  
...J_{m_{n-2}-m_{n-1}}\left(\frac{kK}{s}-(m_{1}+m_{2}+...+m_{n-2})   
K \right)  
\exp \left(-\frac{\sigma^{2}}{2}m_{n-2}^{2} \right)  
\]  
\[  
J_{m_{n-1}-0}\left(\frac{kK}{s}\right)  
\exp \left(-\frac{\sigma^{2}}{2}m_{n-1}^{2}\right)  
\delta_{m_{1}+m_{2}+...+m_{n-2}+m_{n-1},0}.  
\]  
Thus in order to obtain the expansion coefficient $a_{n}$ we should perform  
summation in (\ref{u14}) over all integers  
subject to the constraint  
\begin{equation}  
m_{1}+m_{2}+...+m_{n-1}=0.  
\label{u15}  
\end{equation}


  

We are interested in the limit  
of large $s$ and $K$. The limits are taken in the order  
\begin{equation}  
(1)\  s \rightarrow \infty,\   
\label{u16}  
(2)\ K \rightarrow \infty,\  
(3)\  \sigma \rightarrow 0.  
\end{equation}  
Finite $\sigma$ is required to assure the absolute   
convergence of the series.  
Therefore (\ref{u14}) is summed for finite $\sigma$ and  
the limit   
$\sigma \rightarrow 0$ should be taken   
in the end of the calculation.  
Having this limit in   
mind the leading term in $K /s$ and $1/ \sqrt{K}$ will be identified.  
It will be assumed that the mode that is calculated  
is sufficiently low so that   
\begin{equation}  
0<k K/s<<1.  
\label{ad2}  
\end{equation}  
Each term in (\ref{u14}) is defined by  the string  
\[  
(m_{1},m_{2},...,m_{i},...,m_{n-1}).  
\]  
The leading contribution  
\begin{equation}  
a_{n}^{(0)}=J_{0}^{n} \left(\frac{kK}{s} \right)   
\approx \left(1-\frac{k^{2}K^{2}}{4s^{2}} \right)^{n}  
\label{u17}  
\end{equation}  
results from the string where all $m_{j}$ vanish. A nonvanishing  
$m_{j}$ results in a Bessel function with a large argument, since  
$K /s<<1$ and $K>>1$, and therefore it leads to  
a factor $1/ \sqrt{K}$ in the contribution to $a_{n}$.   
Let $m_{i}$ be the first   
nonvanishing $m_{j}$ and $m_{f}$ the last nonvanishing one. The first factor   
in (\ref{u14}) that is not $J_{0}(\frac{kK}{s})$ is   
$J_{-m_{i}} \left(\frac{kK}{s} \right)$  
and the last factor that differs from $J_{0}(\frac{kK}{s})$ is   
  
\begin{equation}  
J_{m_{f}} \left(\frac{kK}{s}-(m_{1}+m_{2}+...+m_{f})K \right)  
=J_{m_{f}} \left(\frac{kK}{s} \right).  
\label{e2}  
\end{equation}  
Since  
$J_{n}(x)\approx\frac{x^{n}}{2^{n}n!}$   
for small $x$ and  
$J_{-n}(x)=(-1)^{n}J_{n}(x)$  
the contribution of the terms between $i$ and $f$ is of the order   
  
\begin{equation}  
C \ \left(\frac{kK}{s} \right)^{|m_{i}|+|m_{f}|}\ ,  
\label{u18}  
\end{equation}  
where $C$ is the contribution of the factors with $m_{j}$ that are not the first  
and last ones.  
The first factor after $m_{i}$ is    
\[  
J_{m_{i}-m_{i+1}} \left(\frac{kK}{s}-(m_{1}+...+m_{i})K \right)=  
J_{m_{i}-m_{i+1}} \left(\frac{kK}{s}-m_{i}K \right)  
\]   
and the last factor before $m_{f}$ is   
\[  
J_{m_{f-1}-m_{f}}   
\left(\frac{kK}{s}-(m_{1}+...+m_{f-1})K \right)=  
J_{m_{f-1}-m_{f}}  
\left(\frac{kK}{s}-(m_{i}+...+m_{f-1})K \right).  
\]  
The terms in between are of the form    
$J_{m_{j-1}-m_{j}} \left(\frac{kK}{s}-M_{j}K \right)$,  
where $M_{j}=m_{i}+m_{i+1}+...m_{j}\neq 0$ that are of the order   
$1/ \sqrt{K}$.  
Therefore the largest contribution from a string $m_{i},m_{i+1},...,m_{f}$ is  
from the shortest string, namely $f=i+1$. Because of (\ref{u15})   
$m_{i}=- m_{f}$ and because of  
(\ref{u18}) the leading contribution is from the string $m_{i}=-m_{f}=\pm 1$.  
The resulting contribution is  
\begin{equation}  
C=J_{m_{i}-m_{f}} \left(\frac{kK}{s}-m_{i}K \right)=  
J_{\pm2} \left(\frac{kK}{s}-(\pm 1)K \right)   
\approx J_{2}(K).  
\label{u19}  
\end{equation}  
The string can start at $n-2$ places, therefore the leading   
correction to $a_{n}^{(0)}$ is:  
\[  
a_{n}^{(1)}=2(n-2)J_{0}^{n-3} \left(\frac{kK}{s} \right)     
 J_{2} \left(\frac{kK}{s}-K \right)  
J_{1}^{2} \left(\frac{kK}{s} \right)e^{-\sigma^{2}},  
\]  
that is approximated as,  
  
\begin{equation}  
a_{n}^{(1)}\approx2(n-2) \left(1-\frac{k^{2}K^{2}}{4s^{2}} \right)^{n-3}  
J_{2}(K)  
\left(\frac{kK}{2s} \right)^{2}e^{-\sigma^{2}}.  
\label{u20}  
\end{equation}  
The sum of the contributions (\ref{u17}) and (\ref{u20}) is  
\begin{equation}  
a_{n}^{(0)}+a_{n}^{(1)}\sim   
\left(1-\frac{k^{2}K^{2}}{4s^{2}} \right)^{n}+  
2n \left(1-\frac{k^{2}K^{2}}{4s^{2}} \right)^{n-1}J_{2}(K)  
\left(\frac{kK}{2s} \right)^{2}e^{-\sigma^{2}}  
\left(\frac{n-2}{n}\right)\left[\frac{1}{1-\frac{k^{2}K^{2}}{4s^{2}}} \right]^2.  
\label{u21}  
\end{equation}  
In the leading order $\frac{1}{1-\frac{k^{2}K^{2}}{4s^{2}}}\approx 1$   
and \[ \lim_{n \rightarrow \infty}\frac{n-2}{n}=1. \]   
Therefore in the leading order  
\begin{equation}  
a_{n}^{(0)}+a_{n}^{(1)}\sim   
\left[ \left(1-\frac{k^{2}K^{2}}{4s^{2}} \right)+  
2J_{2}(K) \left(\frac{kK}{2s} \right)^{2}e^{-\sigma^{2}}   
\right]^{n}=  
\left[1-\frac{k^{2}K^{2}}{4s^{2}}\left(1-2J_{2}(K)e^{-\sigma^{2}}\right)   
\right]^{n}.  
\label{u22}  
\end{equation}  
The resonance closest to the unit circle, $z_{k}$ is identified   
from (\ref{u10}) as the inverse of the radius of convergence  
\begin{equation}  
z_{k}=1-\frac{k^{2}K^{2}}{4s^{2}}+\frac{k^{2}K^{2}}{4s^{2}}  
2J_{2}(K)  
e^{-\sigma^{2}},  
\label{u23}  
\end{equation}  
or within this order of the calculation as  
\begin{equation}  
z_{k}=exp\left(-\frac{k^{2}K^{2}}{4s^{2}}(1-2J_{2}(K)e^{-\sigma^{2}}) \right).  
\label{u24}  
\end{equation}  
These are the eigenvalues of the diffusion operator with the diffusion   
coefficient  
\begin{equation}  
D(K)=\frac{K^{2}}{4}(1-2J_{2}(K)e^{-\sigma^{2}}),  
\label{u25}  
\end{equation}  
in agreement with the earlier results ~\cite{RW}.

The analysis of the off-diagonal matrix elements  
\begin{equation}  
a_{n}=(km|\hat{U}^{n}|k'm')  
\label{ad1}  
\end{equation}  
is similar.  
We assume again (\ref{ad2}) and $K>>1$.  
Analogously to (\ref{u14}) one obtains  
  
\begin{equation}  
a_{n}=\sum_{m_{1}}\sum_{m_{2}}...  
\sum_{m_{n-1}}  
J_{m-m_{1}}\left(\frac{kK}{s}-mK \right)\exp\left(-\frac{\sigma^{2}}{2}m^{2} \right)  
\label{ad3}  
\end{equation}  
\[  
J_{m_{1}-m_{2}}\left(\frac{kK}{s}-(m+m_{1})K \right)  
\exp\left(-\frac{\sigma^{2}}{2}m_{1}^{2} \right)  
J_{m_{2}-m_{3}}\left(\frac{kK}{s}-(m+m_{1}+m_{2})K \right)  
\exp\left(-\frac{\sigma^{2}}{2}m_{2}^{2} \right)  
\]  
\[  
...J_{m_{n-2}-m_{n-1}}  
\left(\frac{kK}{s}-(m+m_{1}+m_{2}+...+m_{n-2})K \right)  
\exp\left(-\frac{\sigma^{2}}{2}m_{n-2}^{2} \right)  
\]  
\[  
J_{m_{n-1}-m'}\left(\frac{k'K}{s} \right)  
\exp\left(-\frac{\sigma^{2}}{2}m_{n-1}^{2} \right)  
\delta_{(m+m_{1}+m_{2}+...+m_{n-2}+m_{n-1})s,k-k'}  
\]  
Because of the last $\delta-$ function  
the $a_{n}\neq 0$ only if $(k-k')/s=q$ is integer.  
The leading contribution results from the string  
$m_{1}=-m$, $m_{n-1}=q$ and all other $m_{j}$ vanish.  
It is therefore of the form  
\begin{equation}  
a_{n}^{(0)}=B\ J_{0}^{n-4}\left(\frac{kK}{s} \right),  
\end{equation}  
where  
\begin{equation}  
B=J_{2m}(mK)J_{-m}\left(\frac{kK}{s} \right)J_{-q}\left(\frac{kK}{s} \right)  
J_{q-m'}\left(\frac{k'K}{s} \right)\exp(-\sigma^{2} m^{2})  
\exp\left(-\frac{\sigma^{2}}{2}q^{2} \right)  
\end{equation}  
that behaves as $a_{n}^{(0)}$ of (\ref{u17}) in   
the large $n$ limit. The leading correction is found from   
neighboring pairs $m_{i}=-m_{i+1}= \pm 1$ as  
in the case studied before with a result similar to  
$a_{n}^{(1)}$ of (\ref{u20}) in the large $n$ limit.  
Therefore no new resonances are found from the off-diagonal  
terms with $k \neq 0$, in the order of approximation that was used.

For $s \gg 1$ the diffusion modes in momentum space constitute  
the slow degrees of freedom of the system. However, the faster relaxation  
modes (or, alternatively the modes of a system with $s \sim 1$)  
cease to be angle independent.  
To evaluate the magnitude of such a fast relaxation rate within our 
perturbation scheme, we have to   
calculate matrix elements associated with  
relaxation of disturbances from the invariant density that involve functions 
from  
the angular subspace $\{ \ |0,m)\ \} $ with $m \neq 0$.  
Consider, therefore, the matrix element 
\begin{equation}  
a_{n}=(0m|\hat{U}^{n}|km').  
\label{w2}  
\end{equation}  
The equation corresponding to (\ref{u13}) is:  
\begin{equation}  
a_{n}=  
\sum_{k_{1},m_{1}}\sum_{k_{2},m_{2}}...\sum_{k_{n-1},m_{n-1}}  
(0m|\hat{U}|k_{1}m_{1})(k_{1}m_{1}|\hat{U}|k_{2}m_{2})...  
(k_{n-1}m_{n-1}|\hat{U}|km')=  
\label{w3}  
\end{equation}  
\[  
\sum_{k_{1},m_{1}}\sum_{k_{2},m_{2}}...  
\sum_{k_{n-1},m_{n-1}}  
J_{m-m_{1}}\left(\frac{k_{1}K}{s} \right)  
\exp\left(-\frac{\sigma^{2}}{2}m^{2} \right)\delta_{-k_{1},ms}  
J_{m_{1}-m_{2}}\left(\frac{k_{2}K}{s} \right)  
\exp\left(-\frac{\sigma^{2}}{2}m_{1}^{2} \right)\delta_{k_{1}-k{2},m_{1}s}  
\]  
\[  
J_{m_{2}-m_{3}}\left(\frac{k_{3}K}{s} \right)  
\exp\left(-\frac{\sigma^{2}}{2}m_{2}^{2} \right)\delta_{k_{2}-k{3},m_{2}s}  
...J_{m_{n-1}-m'}\left(\frac{kK}{s} \right)  
\exp\left(-\frac{\sigma^{2}}{2}m_{n-1}^{2} \right)\delta_{k_{n-1}-k,m_{n-1}s}  
\]  
and summation over the $k_{i}$ yields a nonvanishing result only if  
$k/s \equiv q$ that is an integer. In this case:  
\begin{equation}  
a_{n}=\sum_{m_{1}}\sum_{m_{2}}...  
\sum_{m_{n-1}}  
J_{m-m_{1}}(-mK)\exp\left(-\frac{\sigma^{2}}{2}m^{2} \right)  
\label{w5}  
J_{m_{1}-m_{2}}((-m-m_{1})K)  
\exp\left(-\frac{\sigma^{2}}{2}m_{1}^{2} \right)  
\end{equation}  
\[  
J_{m_{2}-m_{3}}((-m-m_{1}-m_{2})K)  
\exp\left(-\frac{\sigma^{2}}{2}m_{2}^{2} \right)  
...J_{m_{n-2}-m_{n-1}}((-m-m_{1}-m_{2}-...-m_{n-2})K)  
\exp\left(-\frac{\sigma^{2}}{2}m_{n-2}^{2} \right)  
\]  
\[  
J_{m_{n-1}-m'}(q K)  
\exp\left(-\frac{\sigma^{2}}{2}m_{n-1}^{2} \right)  
\delta_{-m-m_{1}-m_{2}-...-m_{n-2}-m_{n-1},q}.  
\]  
The result is independent of $s$.  
This is a sum over $m_{i}$ constrained by  
\begin{equation}  
m+m_{1}+m_{2}+...+m_{n-2}+m_{n-1}=-q.  
\label{c5}  
\end{equation}  
In every particular term in this multiple series, generally, we will have  
multiples of terms $J_{\nu}(MK)$.  
If $M=0$ and $\nu \neq 0$, such a term vanishes while if both M  
and $\nu$ do not vanish $J_{\nu}(MK) \sim \frac{1}{\sqrt{K}}$.  
The leading contribution is from sequences with the maximal   
number of factors $J_{0}(0)=1$. To identify these, we denote  
$J_{0}(0)=1$ by ``1'' and other factors by ``$x$''.  
In this way, to every term in (\ref{w5}) corresponds  
the sequence of $n$ symbols:  
\begin{equation}  
x*x*1*x*x*x*1*x*1*x*x*...*1*x*x.  
\label{w6}  
\end{equation}  
A crucial restriction is that if $m \neq 0$, two ``1''   
symbols cannot be nearest neighbors  
as is shown in what follows.  
If $m \neq 0$ the sequence starts with ``$x$'' as is clear from    
(\ref{w5}). Let the $i$-th symbol be ``1''. Then  
\begin{equation}  
J_{m_{i-1}-m_{i}}((-m_{1}-m_{2}-...-m_{i-1})K)= J_{0}(0).  
\label{w7}  
\end{equation}  
The previous term is  
\begin{equation}  
J_{m_{i-2}-m_{i-1}}((-m_{1}-m_{2}-...-m_{i-2})K).  
\label{w8}  
\end{equation}  
For both to be $J_{0}(0)$ it is required that $m_{i-1}=0$,   
and $m_{i-2}-m_{i-1}=0$, implying $m_{i-2}=0$ resulting in   
\[  
J_{m_{i-3}}((-m_{1}-m_{2}-...-m_{i-3})K)=0,  
\]  
for $m_{i-3} \neq 0$. Therefore if the term before the $i$-th   
one is ``1'' ( and we have two neighbors that are ``1''s )  
either $m_{i-3}$, and all $m_{j}$ with $j < i-2$,  
vanish and all factors before the $i$-th are ``1''s,  
in contradiction with the fact that for $m \neq 0$ the sequence starts   
with an $x$, or the contribution of the sequence   
vanishes ( when one of the $m_{j}$ does not vanish ). Now one has to find  
the strings (\ref{w6}) with the maximal number of ones subject to given   
values of $m$, $m'$ and $q$. For this purpose strings with alternating  
``$x$'' and ``1'' are constructed.  
 
The ``$x$'' represent factors $J_{m_{l}-m_{l+1}}(-M_lK)$ where $M_l=m+\sum_{i=1}^l m_i$ and we  
have to choose the $m_i$ so that the $J_{m_{l}-m_{l+1}}(-M_lK)$ are of maximal  magnitude.  
Consider the string:  
\begin{equation}  
.....J_{m_{l}-m_{l+1}}(-M_lK)~J_{m_{l+1}-m_{l+2}}(-M_{l+1}K)~J_{m_{l+2}-m_{l+3}}(-M_{l+2}K)~  
J_{m_{l+3}-m_{l+4}}(-M_{l+3}K).....  
\label{ex1}  
\end{equation}  
where the factors $\exp(-\frac{\sigma^{2}}{2}m_{l}^{2})$ were omitted for the sake of  
brevity. Requiring that the second and fourth factors are ``1'' yields $m_{l+1}=m_{l+2}$ and  
$m_{l+3}=m_{l+4}$ as well as $M_{l+1}=M_{l+3}=0$. Therefore $M_l=-m_{l+1}$ and  
$M_{l+2}=m_{l+2}=-m_{l+3}$. This implies $m_{l+1}=m_{l+2}=-m_{l+3}=-m_{l+4} \equiv m^*$, and  
this string takes the form:  
\begin{equation}  
.....J_{m_{l}-m^*}(m^* K)~J_0(0)~J_{2m^*}(-m^*K)~J_0(0)....  
\label{ex2}  
\end{equation}  
Continuation of the string to the left requires $m_l=-m^*$. The factors ``$x$'' are  
$J_{2m^*}(-m^*K)=J_{-2m^*}(m^*K)=J_{2m^*}(m^*K)$. For each value of $K$ we choose the value of  
$m^*$ so that $\mid J_{2m^*}(m^*K)\mid $ is maximal, namely   
\begin{equation}  
\mid J_{2m^*}(m^*K) \mid =\max_m {\mid J_{2m}(mK)} \mid.  
\label{ex3}  
\end{equation}  
Now it is left to match this string to the ends that are determined by $m$, $m'$ and $q$. The  
term (\ref{w5}) is the sum of terms of the form:  
\begin{equation}  
C^{(l)}(m,m^*)\left( J_{2m^*}(m^*K) \exp(-\sigma^2 m^{* 2}) \right)^{n'} C^{(r)}(m^*,m',q)  
\label{ex4}  
\end{equation}  
where $n'$ is an integer of the order $n/2$. The string (\ref{ex2}) is of period $4$ therefore  
also the end terms are of period $4$ in $n$. 
One can always find enough  
values of $m_i$ in the beginning and in  
the end of strings in (\ref{w5}) so that they take the form (\ref{ex4}).  
The end terms $C^{(l)}(m,m^*)$ and  $C^{(r)}(m^*,m',q)$ are sums of the contributions of 
these $m_i$. Some of the contributions to the end terms are presented in App. B.  
The end terms do not affect the  
resonance. Therefore the largest resonance associated with the fast decaying modes, corresponding to  
the slowest one, is up to the  
fourth root of the identity:  
\begin{equation}  
\tilde{z}=\sqrt{J_{2m^*}(m^*K)~\exp\left(-\frac{\sigma^2}{2}m^{* 2}\right)},  
\label{ex7}  
\end{equation}  
independent of $m$, $m'$ and $q$. The reason that $\tilde{z}$ is determined only up to  
the fourth root of the identity is the period $4$ of the string (\ref{ex2}). The  
resonances associated with the other fast decaying modes cannot be determined in the framework  
of the perturbative expansion of the present work.

The Frobenius-Perron operator is the evolution operator $\hat{U}$ in the limit  
of vanishing noise. Therefore the Ruelle resonances are the poles of    
matrix elements of the  
resolvent $ \hat{R} $ in this limit. They form several groups.  
There is  
\begin{equation}  
 z_{0}=1,  
\end{equation}  
that is related to the equilibrium state found for $m=m'=q=0$. The resonances  
corresponding to the  
relaxation modes related to the diffusion in the angular momentum space
 are given by  
\begin{equation}  
z_{k}=exp\left(-\frac{k^{2}K^{2}}{4s^{2}}(1-2J_{2}(K) )\right).  
\label{resk}  
\end{equation}  
Finally, the largest resonance related to relaxation in  
the $\theta$ direction is, up  
to the fourth   
root of the identity,  
\begin{equation}  
\tilde{z}=\sqrt{J_{2m^*}(m^* K)},   
\label{resp}  
\end{equation}  
where $m^*$ is chosen so that the expression is maximal for a given value of $K$.  
The corresponding relaxation rates $ \gamma_{k}$ and $\tilde{\gamma}$ are   
defined by   
\begin{equation}  
z_k=  e^{-\gamma_k}  
\label{gammak1}  
\end{equation}  
and by   
\begin{equation}  
|\tilde{z}|=  e^{-\tilde{\gamma}},   
\label{gammap1}  
\end{equation}  
leading to   
$\gamma_k=|ln z_{k}|$ and $\tilde{\gamma}=|ln |\tilde{z}||$.  
The last resonance may take the four values  
\begin{equation}  
\tilde{z}=\pm  e^{-\tilde{\gamma}},  
\label{gz1}  
\end{equation}  
and   
\begin{equation}  
\tilde{z}=\pm i e^{-\tilde{\gamma}}.  
\label{gz2}  
\end{equation}  
The perturbative calculation enables to compute only $|\tilde{z}|$.    
  
\section{Numerical Exploration of Relaxation}  
  
In Sec. 3 the Ruelle resonances were calculated for large $K$ and 
extrapolated from finite to  
vanishing variance of the noise $\sigma$. Finite noise has the 
effect of truncation of the matrix of  
the FP operator and the limit $\sigma \rightarrow 0$ is the
 infinite matrix limit. In the limit $K  
\rightarrow \infty$ complete stochasticity takes place, 
while for finite $K$ the system is a mixed one,  
but for large $K$ the chaotic component covers nearly all of 
phase space. The results of Sec.  
3 were obtained as the leading terms in an expansion in 
powers of $1/\sqrt{K}$. In the present section  
the results will be tested numerically for finite $K$ and $\sigma=0$.   
The phase space (\ref{d3}) with various  
values of $s$ will be used. The resonances of the type 
(\ref{gammak1}), corresponding to diffusion in angular  
momentum $J$ and of the type (\ref{gammap1}) corresponding to 
relaxation in the $\theta$ direction will be  
calculated numerically from the relaxation rates of various 
perturbations to the  
uniform invariant density. For large $s$, the relaxation of the 
diffusion modes (\ref{gammak1}) (with small $k$) is  
slow and these dominate the long time behavior. To see the angular 
relaxation modes (\ref{gammap1}) one has to  
eliminate the slow relaxation. This can be done either by the 
choice of small $s$ or by the use of  
distributions that are uniform in the momentum $J$. Evolving 
an initial distribution $g$ for $n$ time  
steps and projecting it on a distribution $f$ defines the 
correlation function:  
\begin{equation}  
C_{fg}(n)=(f|\hat{U}^{n}|g).  
\label{cf1}  
\end{equation}  
For a chaotic system, for large $n$ it is expected to decay as   
\begin{equation}  
C_{fg}(n) \sim e^{-\gamma n}  
\label{cfd}  
\end{equation}  
and the relaxation rate $\gamma$ is computed numerically from plots of $C_{fg}(n)$ as a function of  
$n$. In what follows  the distributions $g$ and $f$ will be selected from the Fourier basis (\ref{3}) so  
that $\gamma$ is expected to take the values $\gamma_k$ or $\tilde{\gamma}$.  
Relaxation of the form (\ref{cfd}) is expected to hold in the chaotic   
component. An efficient way to  
calculate correlation functions like (\ref{cf1}) projected on  
this component is from a  
trajectory in phase space.  
By ergodicity it samples all phase space in this component. The phase space integrals involved in the  
calculation of (\ref{cf1}) are replaced by time averages along the trajectory. The trajectories were  
started in the vicinity of the hyperbolic point $(\pi,0)$ and iterated for a large number of time  
steps, $N$.   
It was verified for several cases that the results  
``equilibrize'', namely they do not depend on $N$  
for large $N$.  
The correlation function is calculated from the formula:  
\begin{equation}  
C_{fg}(n)=\lim_{N\rightarrow \infty}\frac{1}{N}\sum_{i=1}^{N}  
f(i)g(i+n)  
\label{cf2}  
\end{equation}  
where $f(j)$ and $g(j)$  are the values of $f$ and $g$ at the $j$-th time step.  
We first calculate numerically the slow relaxation rates $\gamma_k$ (\ref{gammak1}) related to  
diffusion and then turn to calculate   
and $\tilde{\gamma}$ (\ref{gammap1}) 
related to relaxation in the $\theta$  direction.   
  
\subsection{ The Diffusive Modes}  
  
The relaxation rates expected from the approximate calculations of Sec. 3   
for the diffusive  
modes are given by (\ref{resk}) or  
\begin{equation}  
\gamma_k=\frac{k^2}{s^2}D(K),  
\label{gammak2}  
\end{equation}  
where $D(K)$ is the diffusion coefficient (\ref{d1})  
for $\sigma=0$.   
To test this relation, the correlation function  
(\ref{cf1}) was calculated for various distributions $g$ and $f$ from the Fourier basis (\ref{3}) and plots  
like the ones presented in Fig. 1 were prepared. The slope is $\gamma_k$ and the values of $D(K)$ are  
extracted with the help of (\ref{gammak2}) for various values of $k$ and $s$. 
In Fig. 2 these values  
of $D(K)$ are depicted for large values of the stochasticity parameter $K$. 
Excellent agreement with  
the theory is found: (a) The value of $D$ is found to be independent of $k$ and $s$; (b)  
It agrees with the theoretical prediction (\ref{d1}).   
We find indeed that for long time the behavior of distributions is indeed as   
for a diffusive process. In the past it was checked that  
only the second moment of the momentum grows linearly as expected for   
diffusion. The effect of sticking to the accelerator modes was not observed   
for the values of $K$ used for Fig. 2 since the size of the accelerated region   
is small and therefore special effort is required to observe these effects in   
numerical calculations ~\cite{Zaslav}. These are expected to be important for   
relatively small values of $K$ where the accelerated regions are larger.   

In Fig. 3 the correlation function is plotted for relatively small values of   
the stochasticity  
parameter $K$ where   
larger deviations from the theory presented in Sec. 3 are  
expected.  
The diffusion  
coefficient as a function of K is presented in Fig. 4.   
Deviations of the numerical results from the analytical predictions   
are found for some values of $K$. Also for these  
the decay of correlations is found to be exponential and  
the diffusion coefficient extracted for all modes by (\ref{gammak2})   
is the same.  
Therefore the behavior that is found is indeed diffusive, but   
the value of diffusion coefficient for some values of $K$ is larger   
than the one  
that is theoretically predicted. This is a result of sticking ( for  
finite times ) to accelerator modes. For most values of  
$K$ the value of $D$  
found from (\ref{gammak2}) agrees with the one found from direct evaluation   
of trajectories in the chaotic component.  
The theoretical errors (marked by dashed line in Fig. 4 )  
were estimated from the next term of the formula of Rechester   
and White for the diffusion coefficient ~\cite{RW}.

The actual errors are larger   
due to nonperturbative nature   
of the accelerator modes and the surrounding regions   
(such modes cannot be found in an expansion in $ 1/ \sqrt{K}$ ).   
Since in all calculations only trajectories belonging to the chaotic  
component were propagated real acceleration is avoided. The  
trajectories used in the calculation of the correlation function   
by (\ref{cf2})   
effectively generate a projection on the chaotic component of phase  
space.  
\begin{figure}
\begin{center}
\begin{minipage}{7.01cm}
\centerline{\epsfxsize=7.0cm \epsfbox{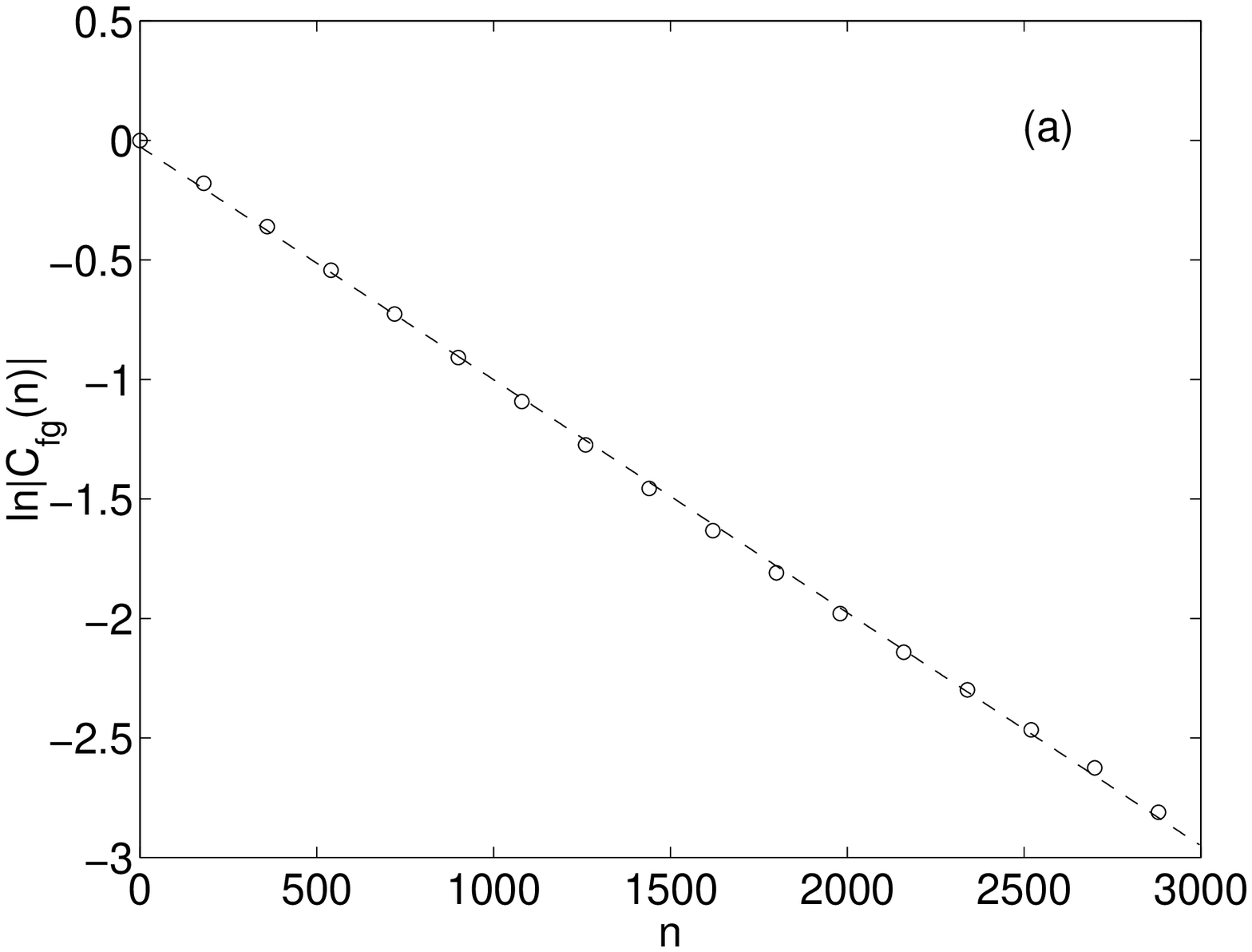} }
\end{minipage}
\hspace{1.0cm}
\begin{minipage}{7.01cm}
\centerline{\epsfxsize=7.0cm\epsfbox{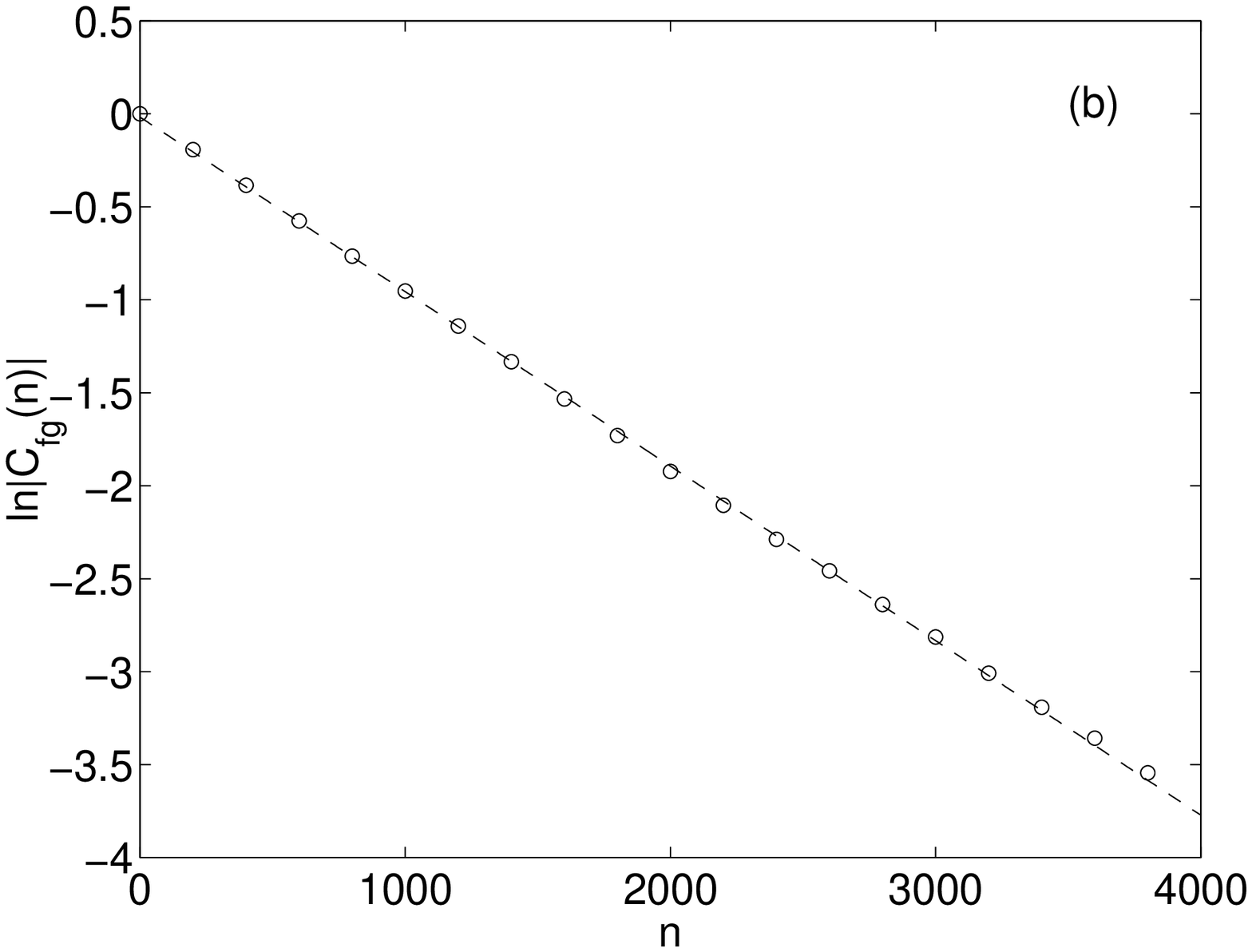} }
\end{minipage}
\\ 
\begin{minipage}{7.01cm}
\centerline{\epsfxsize=7.0cm\epsfbox{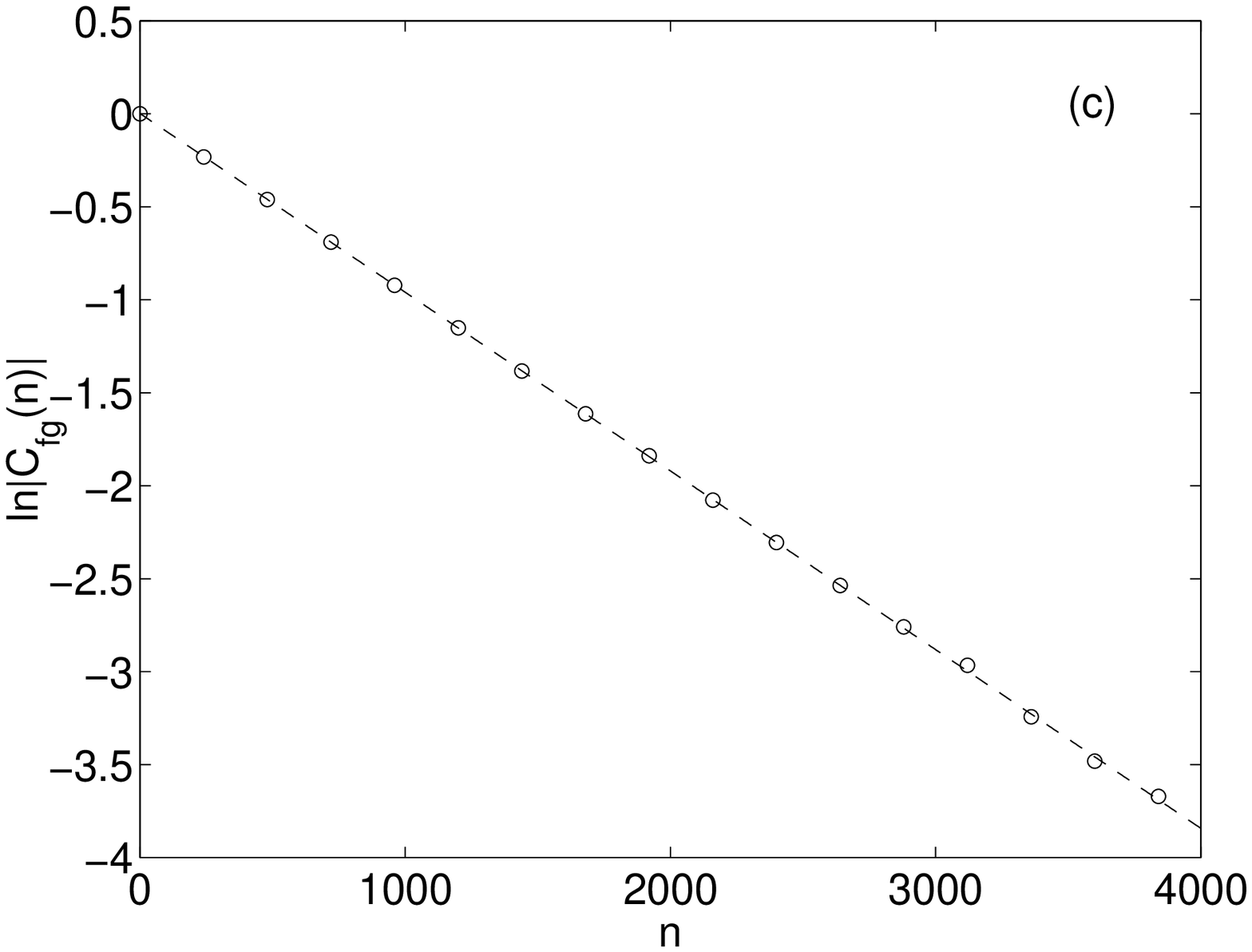} }
\end{minipage}
\hspace{1.0cm}
\begin{minipage}{7.01cm}
\centerline{\epsfxsize=7.0cm\epsfbox{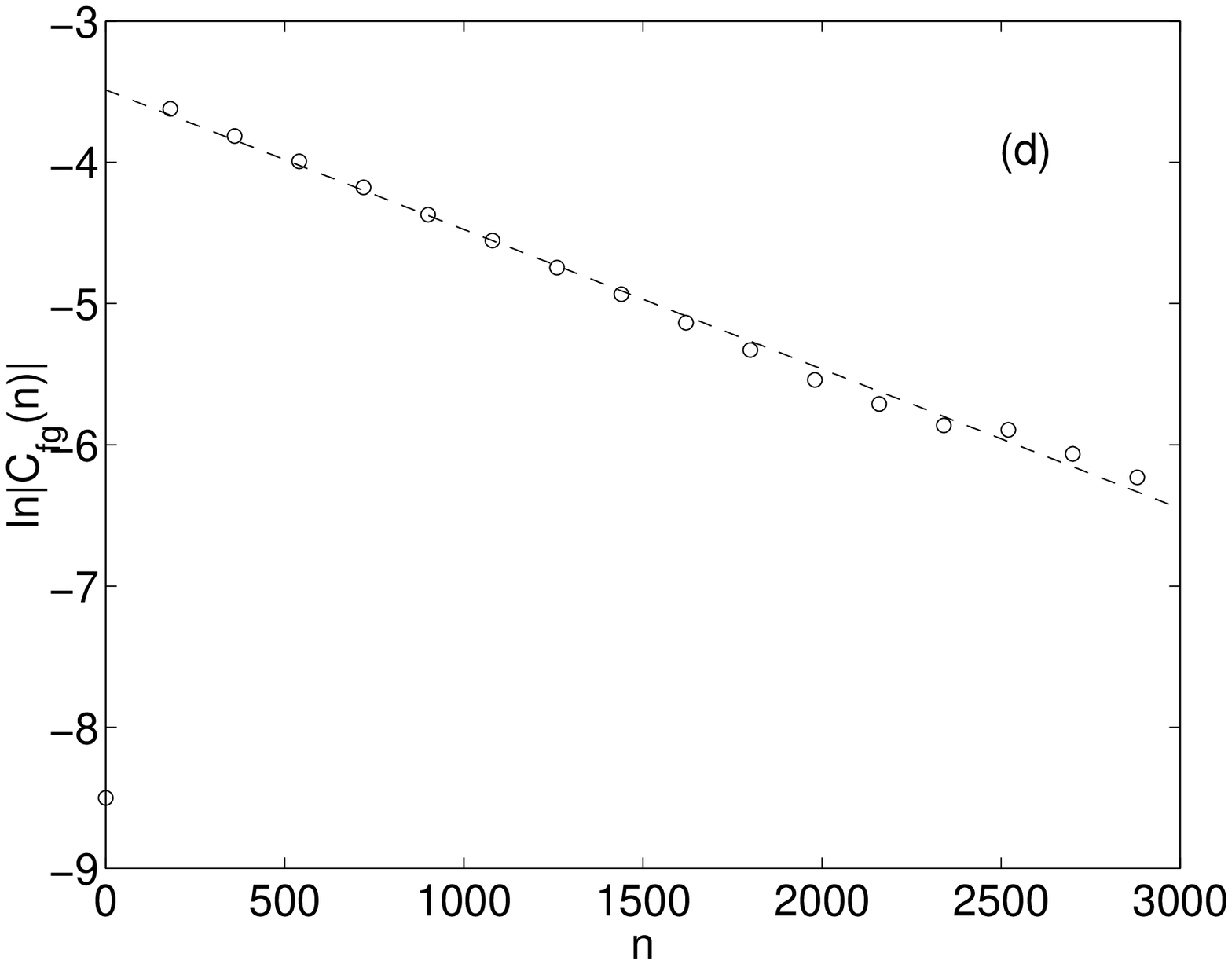} }
\end{minipage}
\end{center}
\caption{The function $C_{fg}(n)$ (semilogarithmic plot) for:
(a) $f=g=\phi_{10}, K=20, s=370, $ ;
(b) $f=g=\phi_{20}, K=30, s=900$;
(c) $f=g=\phi_{50}, K=40, s=3200$;
(d) $f=\phi_{12}, g=\phi_{13}, K=27, s=450$.
The dashed line represents the best fit to the data. 
The number of iterations is $N=8*10^{6}$. }
\end{figure}

\begin{figure}
\centerline{\epsfysize 12.0cm \epsfbox{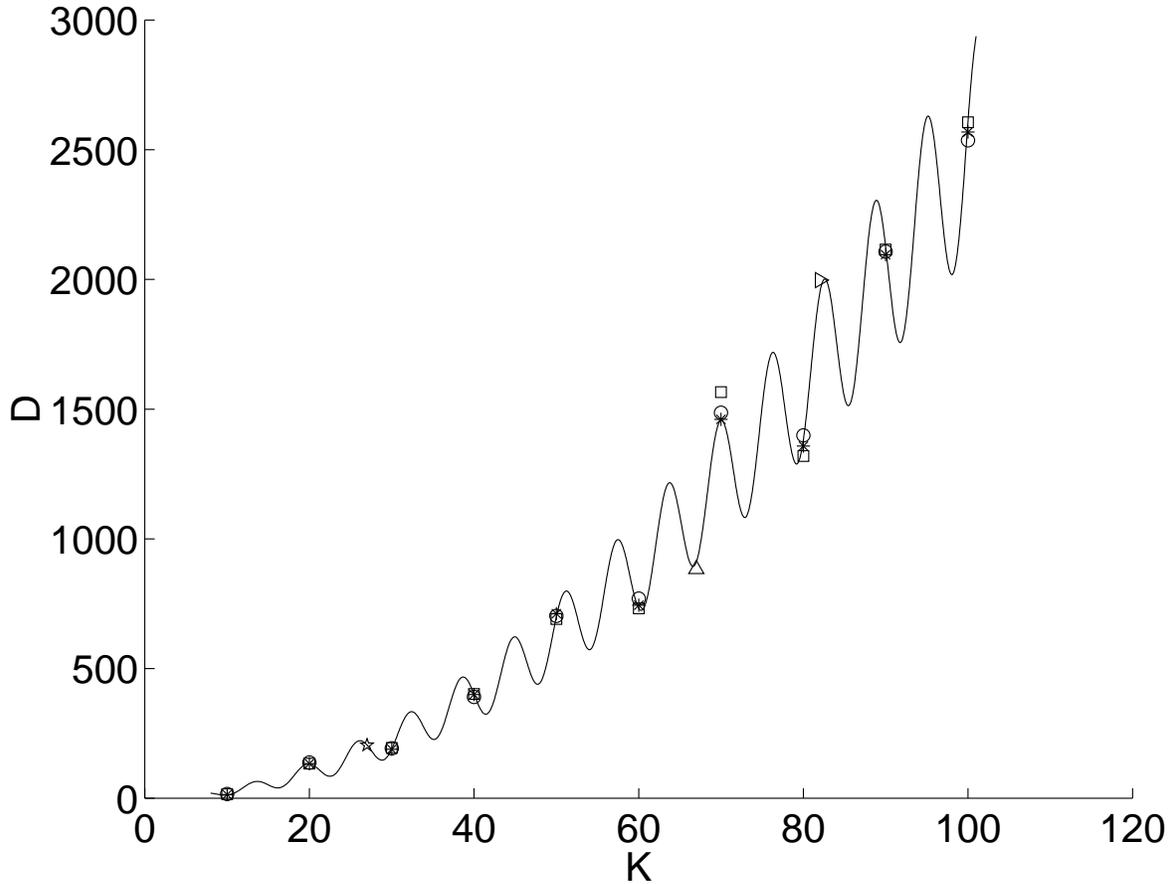}  }
\caption{ The diffusion coefficient $D$ for $K \geq 10$ as found from plots
like  
the ones presented in Fig.1 for: the first mode (1a)
(squares), the second mode (1b) (stars), the fifth mode (1c) 
(circles), correlation function for the
off diagonal first mode (1d) (pentagram)
and other off diagonal correlation functions (triangles),
compared to the
theoretical value (solid line).} 
\end{figure}
\begin{figure}
\begin{center}
\begin{minipage}{7.1cm}
\centerline{\epsfxsize 7.0cm \epsfbox{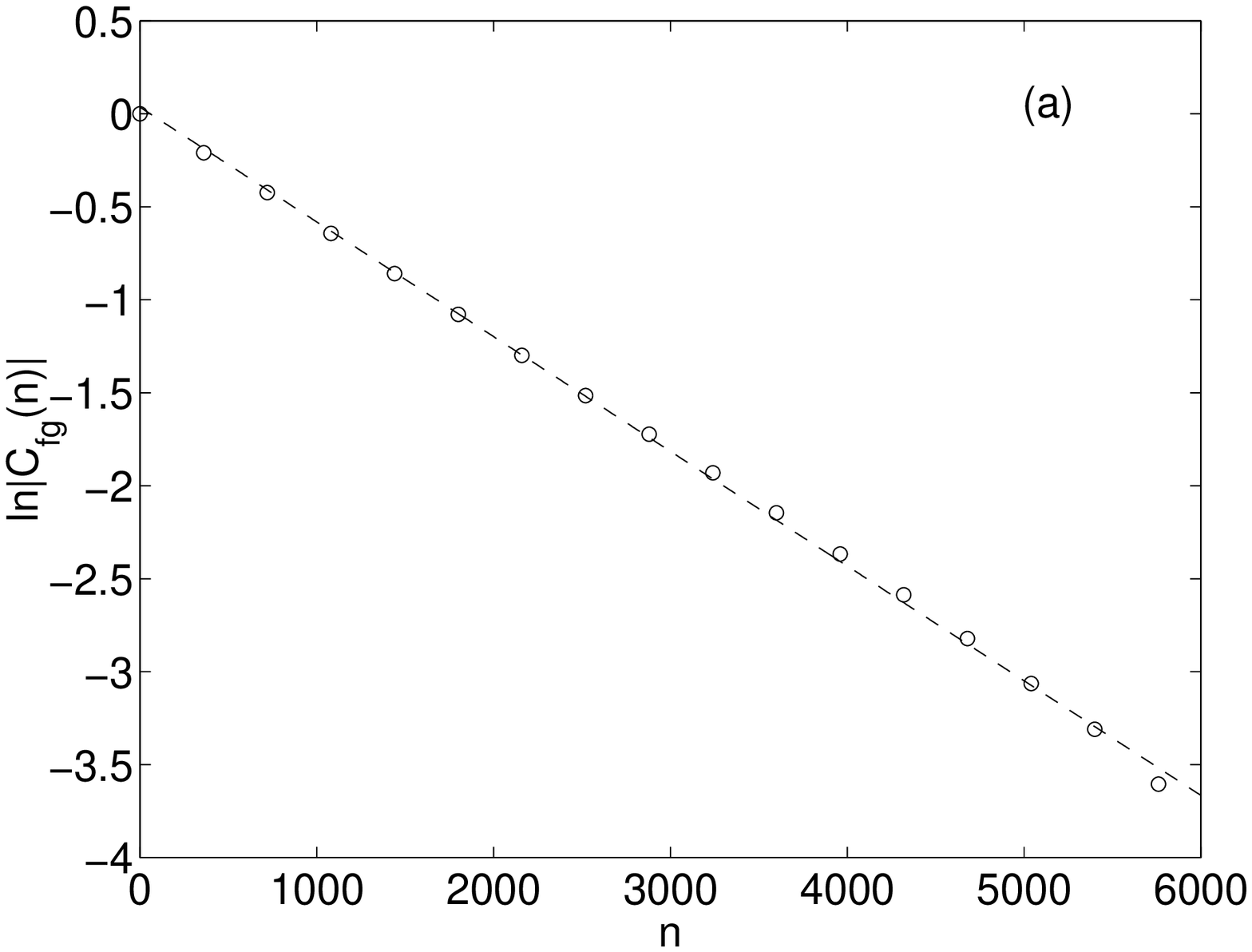} }
\end{minipage}
\hspace{1.0cm}
\begin{minipage}{7.1cm}
\centerline{\epsfxsize 7.0cm\epsfbox{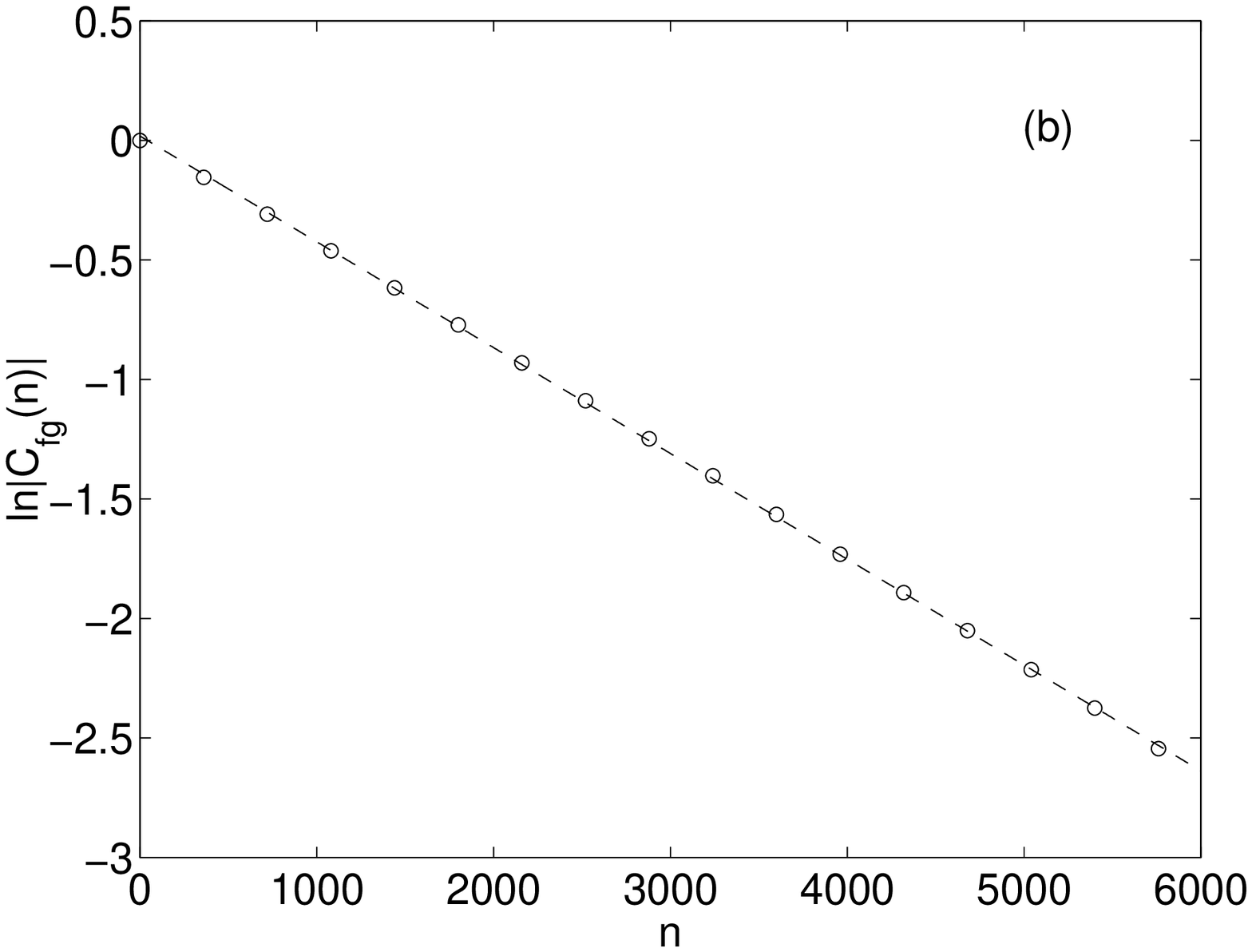} }
\end{minipage}
\\ 
\begin{minipage}{7.1cm}
\centerline{\epsfxsize 7.0cm\epsfbox{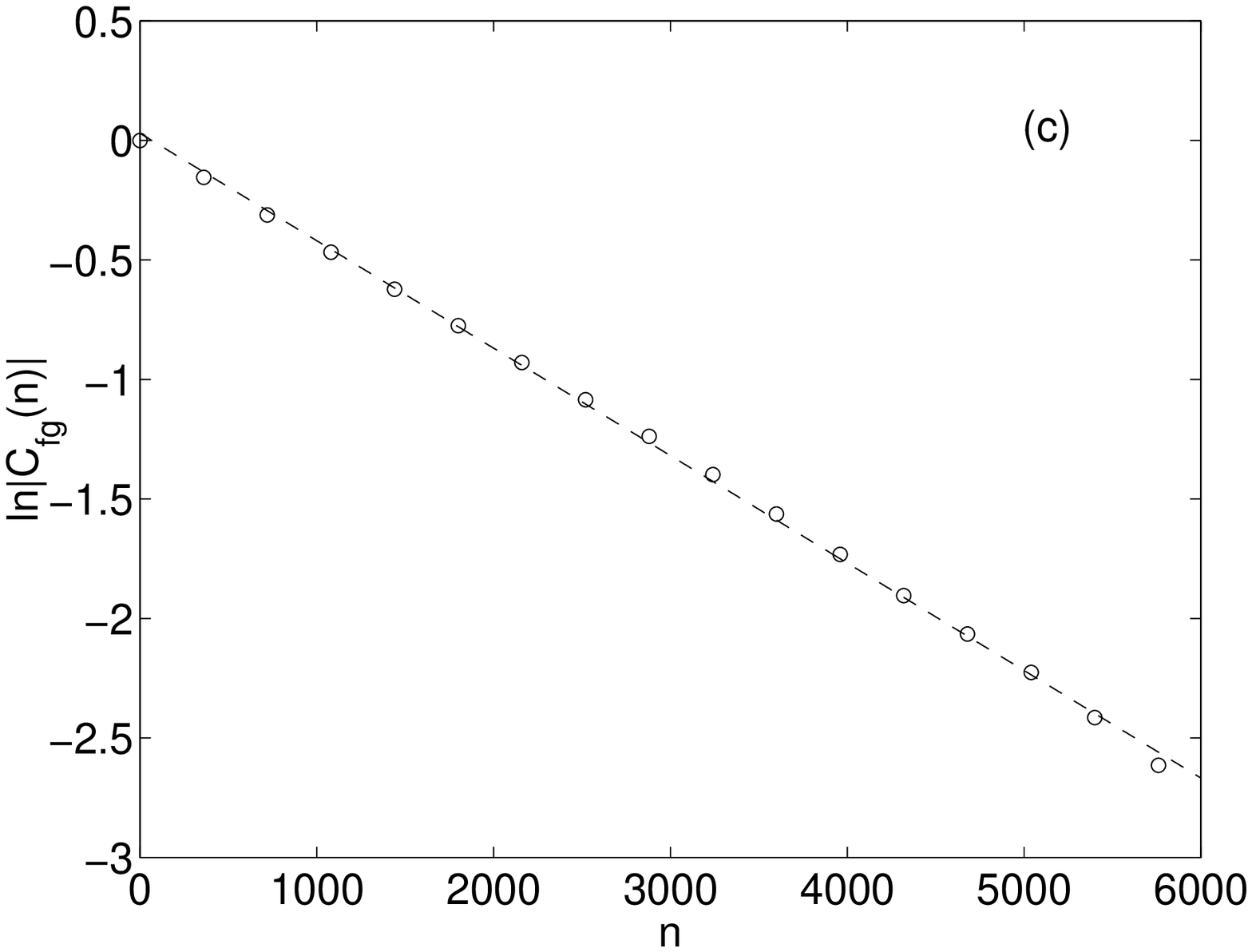} }
\end{minipage}
\hspace{1.0cm}
\begin{minipage}{7.1cm}
\centerline{\epsfxsize 7.0cm\epsfbox{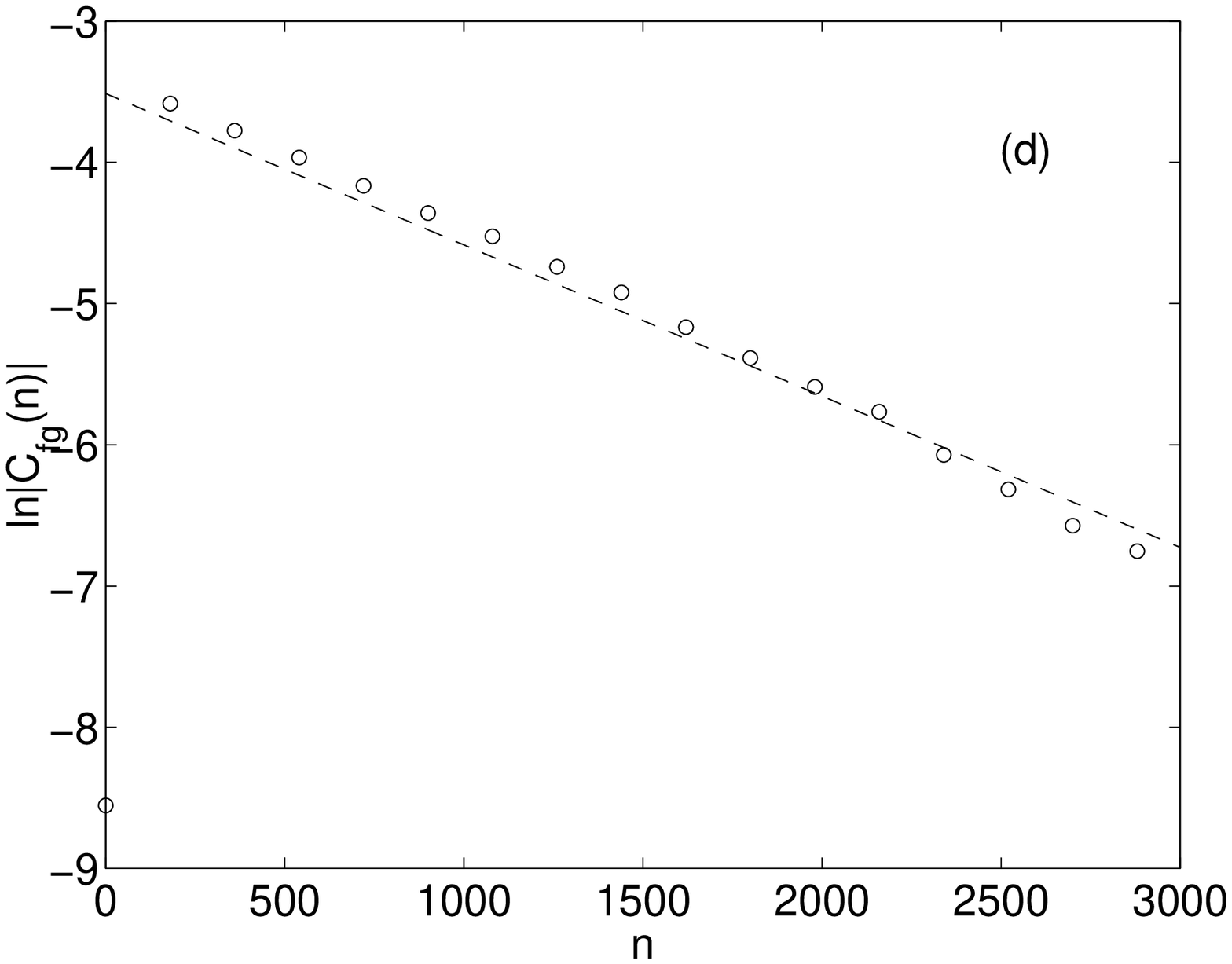} }
\end{minipage}
\hspace{1.0cm}
\end{center}
\caption{The function $C_{fg}(n)$ for 
(a) $f=g=\phi_{10}, K=7, s=250$,
(b) $f=g=\phi_{20}, K=8, s=510$,
(c) $f=g=\phi_{50}, K=3, s=340$,
(d) $f=\phi_{11}, g=\phi_{12}, K=17, s=225$.
The dashed line represents the best fit to the data.
The number of iterations is $N=8*10^{6}$.}
\end{figure}

\begin{figure}
\centerline{\epsfysize 12.0cm \epsfbox{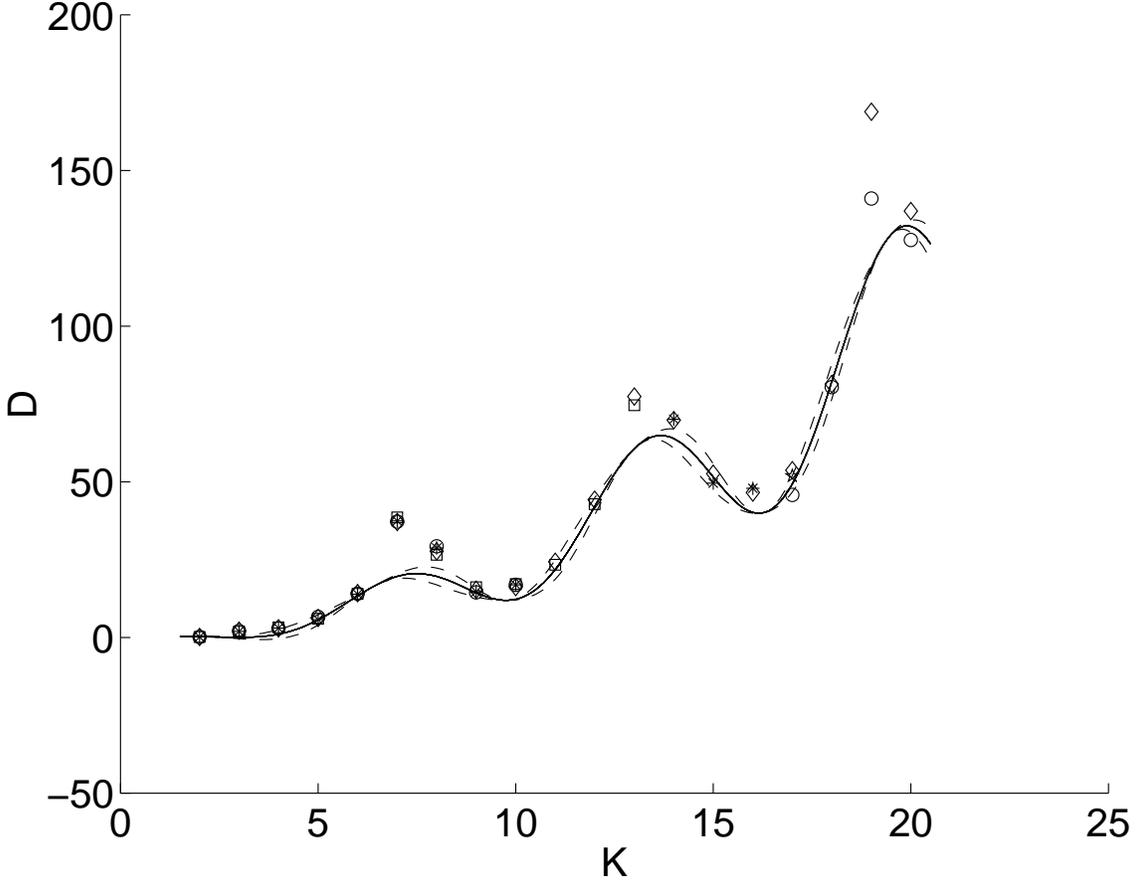}  }
\caption{ The diffusion coefficient $D$ for  $K \leq 20$ as found from plots
like
the ones presented in Fig.3 for: the first mode (1a)
(squares) , the second mode (1b) (stars), the fifth mode (1c)
(circles) and off diagonal first mode (1d) (pentagram), compared to the
theoretical value (solid line).
The dashed line represents the approximate error. The values of $D$ obtained by
direct simulation of propagation of trajectories are marked by diamonds.}
\end{figure}  

\subsection{Angular Relaxation}  
 
In order to observe the angular relaxation mode it is required that no   
relaxation in the $J$ direction is present, because such a relaxation if   
present,   
is expected to dominate the long time limit.   
Since the results are independent on $s$, we use $s=1$.  
For this purpose we take  
$g=\phi_{km'}$ (see (\ref{3})) so that $q=k/s$ is an integer,   
and $f=\phi_{0m}$.  
From (\ref{resp}) and (\ref{gammap1}) one concludes that  
the slowest of the angular relaxation rates is  
\begin{equation}  
\tilde{\gamma}=  
-\frac{1}{2}\ln{[\max_{m^*}\left(|J_{2m^*}(m^*K)| \right)]}.  
\label{gammap2}  
\end{equation}  
The absolute value of the correlation function $ C_{fg}(n)$  
is presented in Figs. 5 and 6  
for $g=\phi_{02}$ and $f=\phi_{01}$ and for $g=f=\phi_{01}$ respectively  
for several values of $K$. The numerical calculations are complicated  
since the relaxation is fast,   
with a characteristic time of the order of one time step.  
Moreover there are oscillations of the correlation function while   
(\ref{gammap2}) is just the envelope. In Figs. 5 and 6 the best fit to the   
envelope is marked by a  
dashed line. The slope of the dashed line is the numerical estimate for   
the relaxation rate. In Fig. 7 the numerical estimate is compared with   
the theoretical prediction.   
The error in the theoretical prediction is estimated as the value of the next   
order contribution to $a_{n}$. This results from a term where the ``1''s   
in sequences corresponding to (\ref{ex2}) and (\ref{ex4}) is replaced by   
an ``$x$'' that represents a   
Bessel function of order $1/\sqrt{K}$,  
leading to an error of the order $\ln{(1\pm 1/\sqrt{K})}$ in the   
relaxation rate. It is difficult to estimate the error resulting from the  
numerical procedure of calculating the relaxation rates. The reason is 
that near the origin of the correlation function a large number of modes 
contributes. On the other hand, in the tail of the correlation function, 
where only one relaxation rate is dominant, the signal is too small. 
Nevertheless, the comparison between our numerical and theoretical results  
shows a good qualitative agreement. 
\begin{figure}
\begin{center}
\begin{minipage}{7.1cm}
\centerline{\epsfxsize 7.0cm \epsfbox{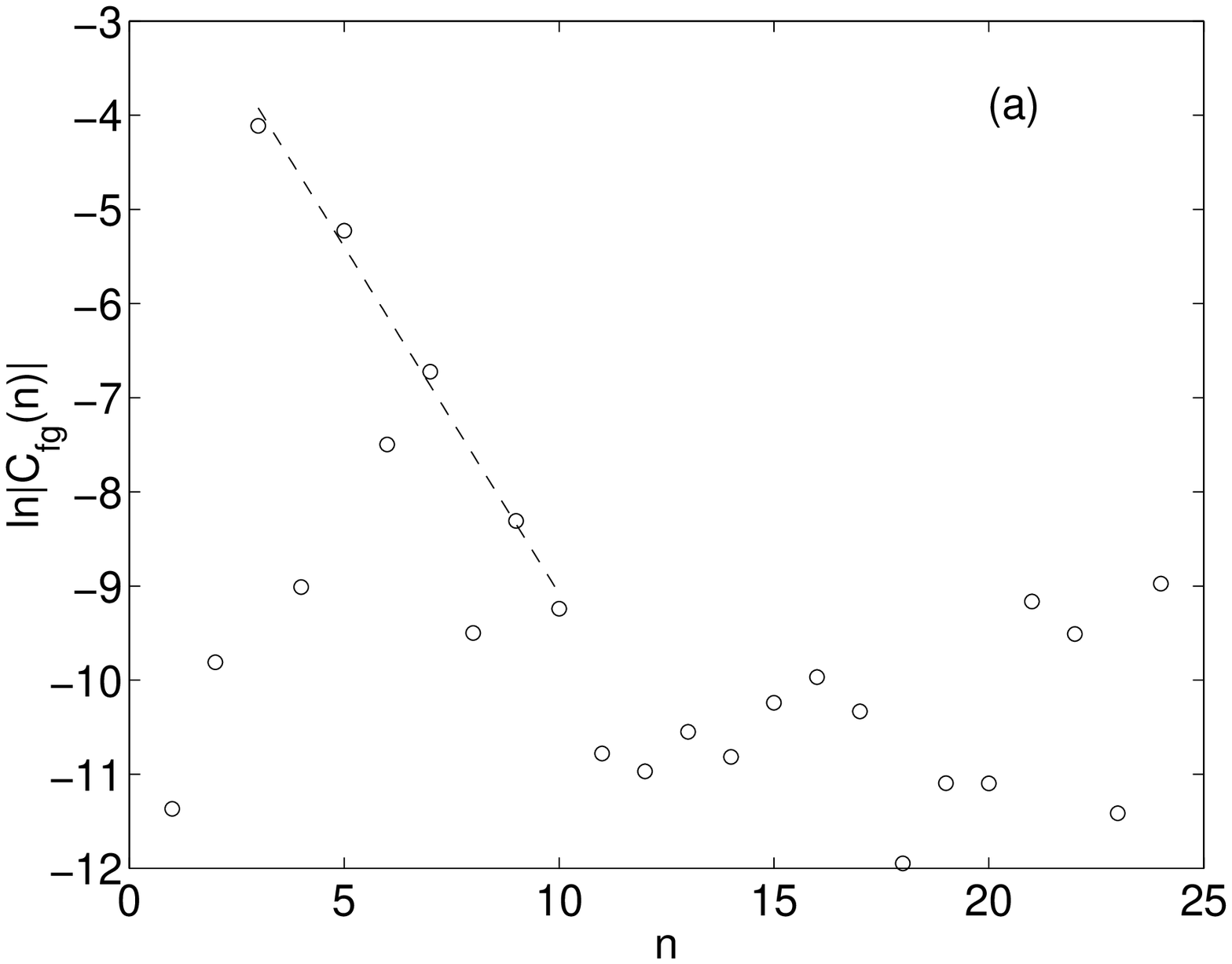} }
\end{minipage}
\hspace{1.0cm}
\begin{minipage}{7.1cm}
\centerline{\epsfxsize 7.0cm\epsfbox{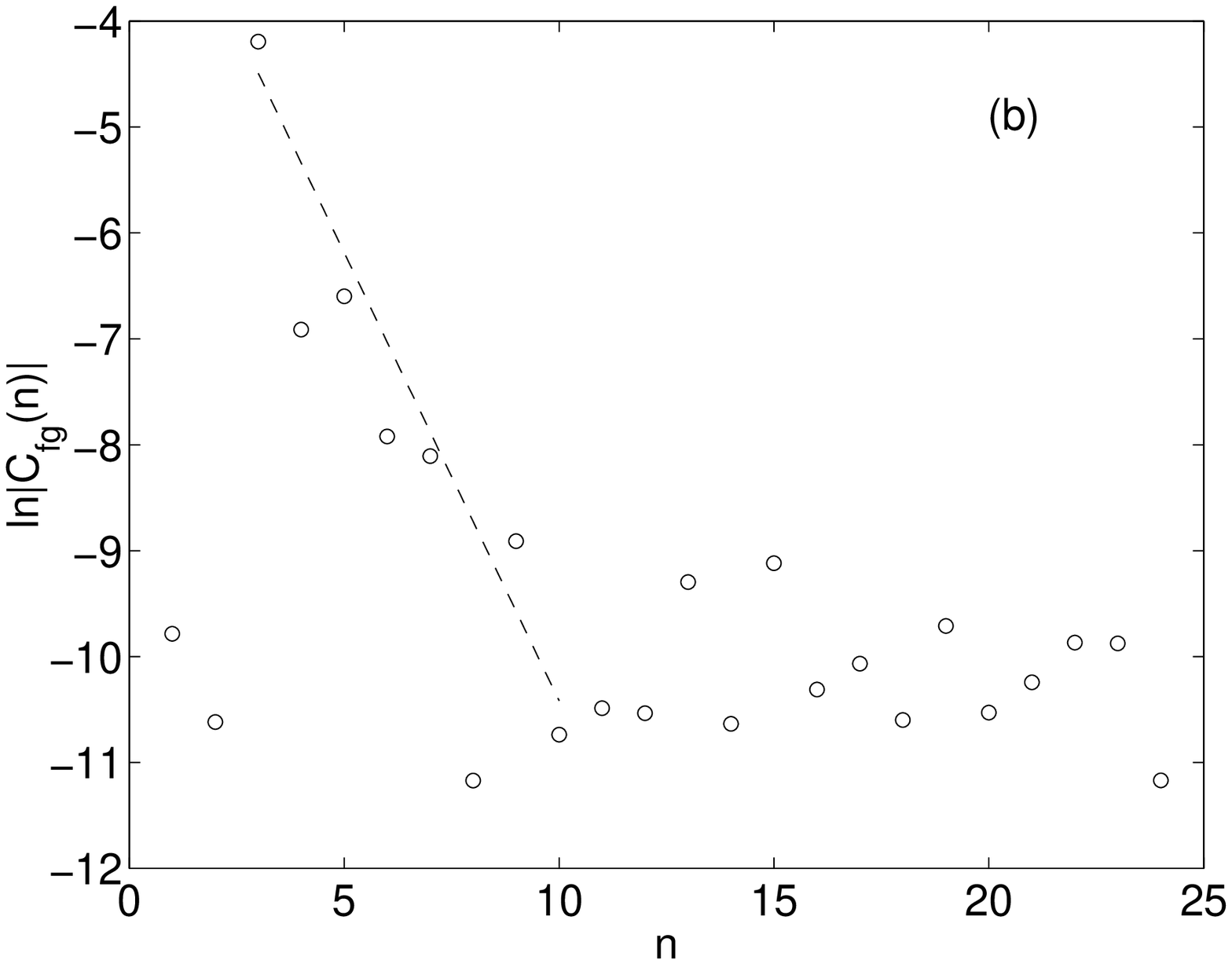} }
\end{minipage}
\\
\begin{minipage}{7.1cm}
\centerline{\epsfxsize 7.0cm\epsfbox{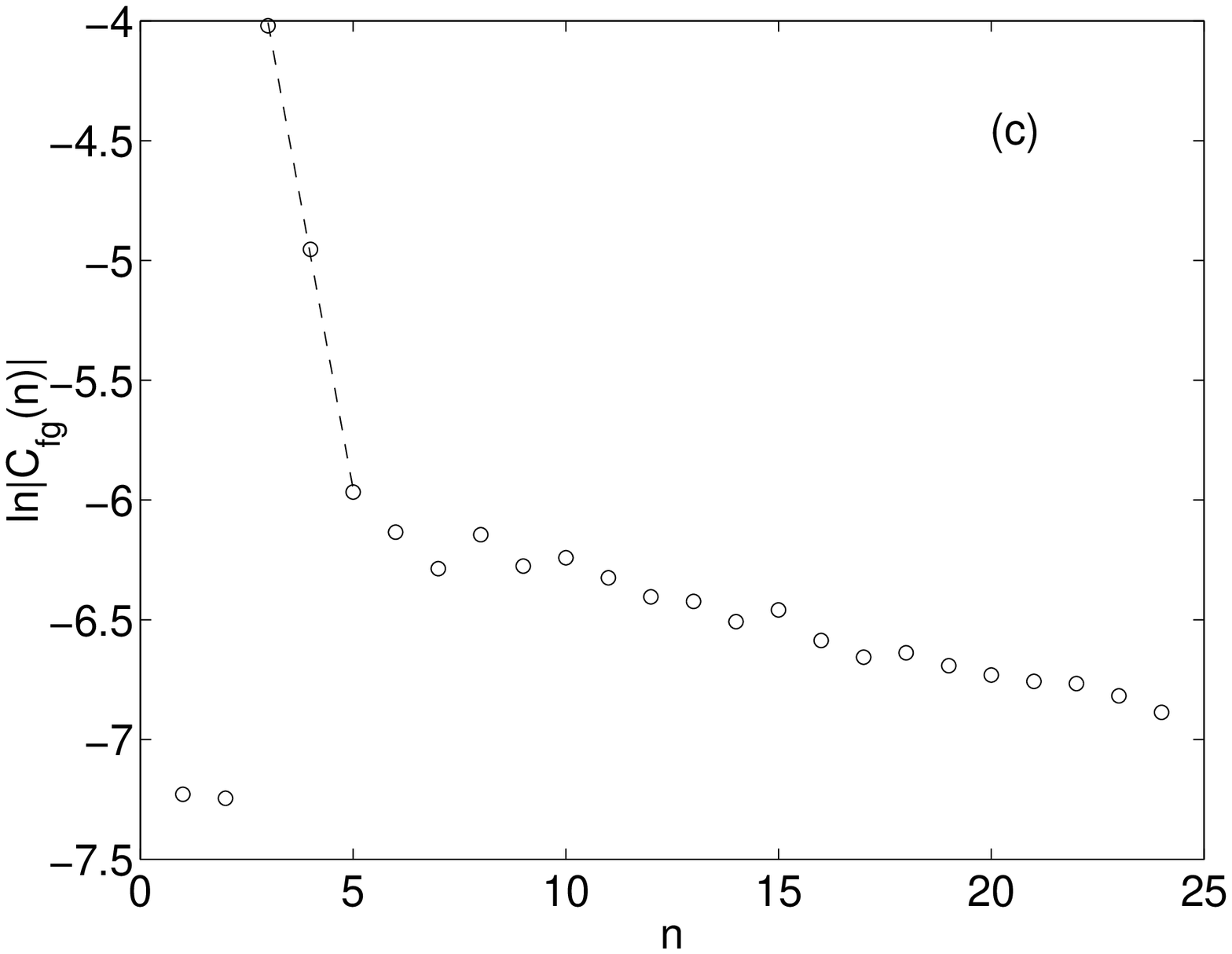} }
\end{minipage}
\hspace{1.0cm}
\begin{minipage}{7.1cm}
\centerline{\epsfxsize 7.0cm\epsfbox{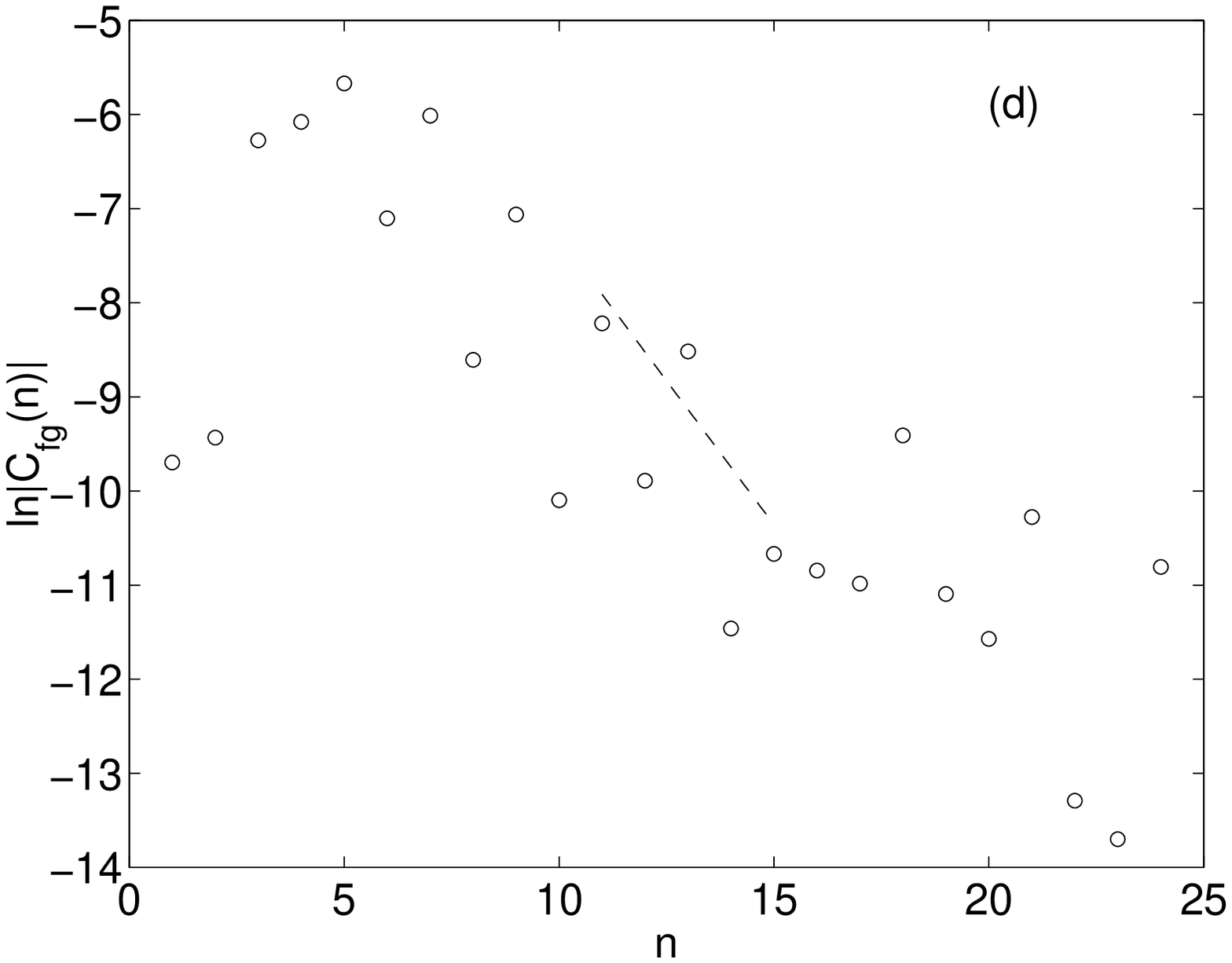} }
\end{minipage}
\hspace{1.0cm}
\end{center}
\caption{The absolute value of the function $C_{fg}(n)$ for 
$f=\phi_{01}, g=\phi_{02} $ and
(a) K=16.3,
(b) K=19.5, 
(c) K=12, 
(d) K=16.
The dashed line represents the best fit to the data.
The values $s=1$ and $N=10^{8}$ were used.}
\end{figure}
\begin{figure}
\begin{center}
\begin{minipage}{7.1cm}
\centerline{\epsfxsize 7.0cm \epsfbox{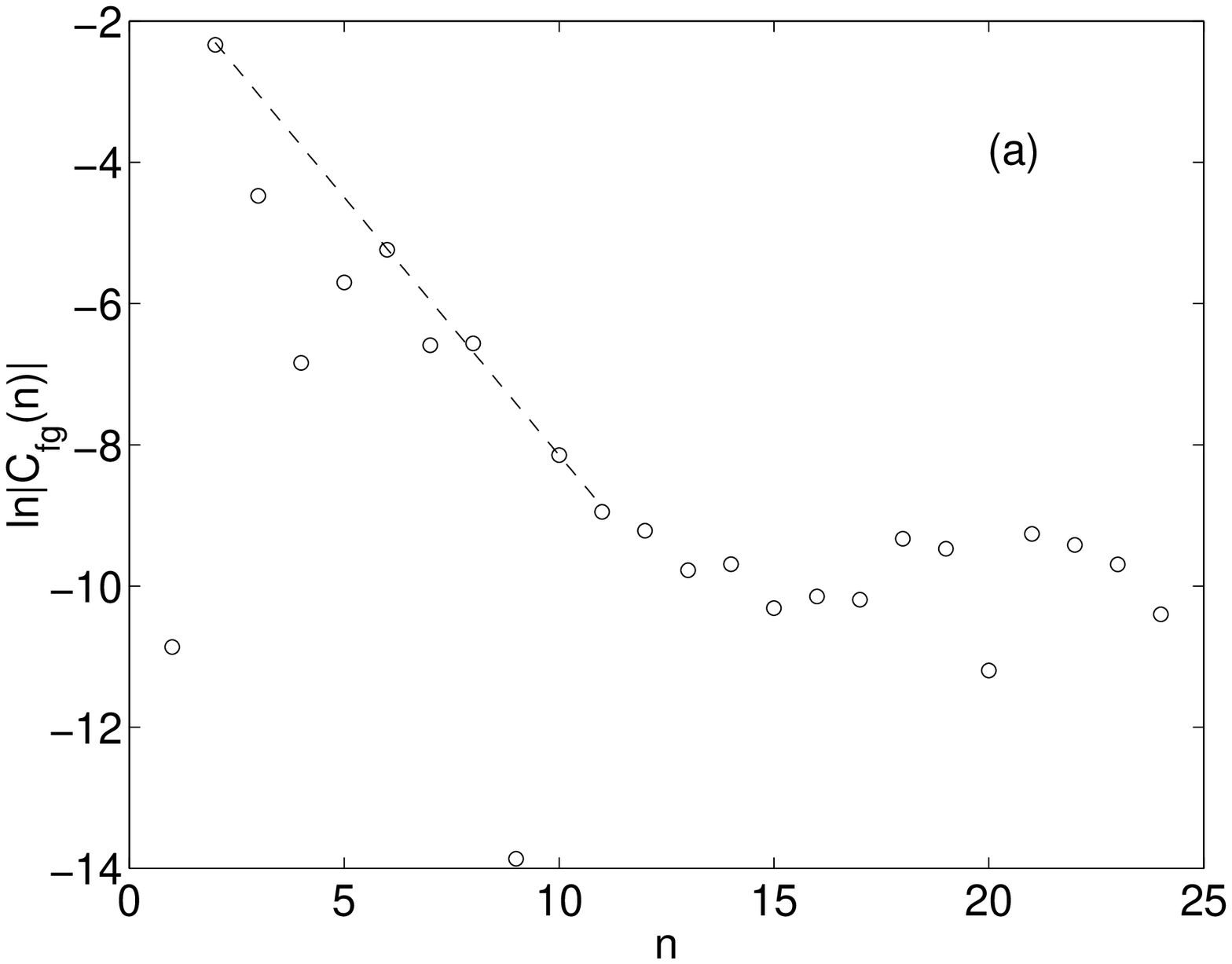} }
\end{minipage}
\hspace{1.0cm}
\begin{minipage}{7.1cm}
\centerline{\epsfxsize 7.0cm\epsfbox{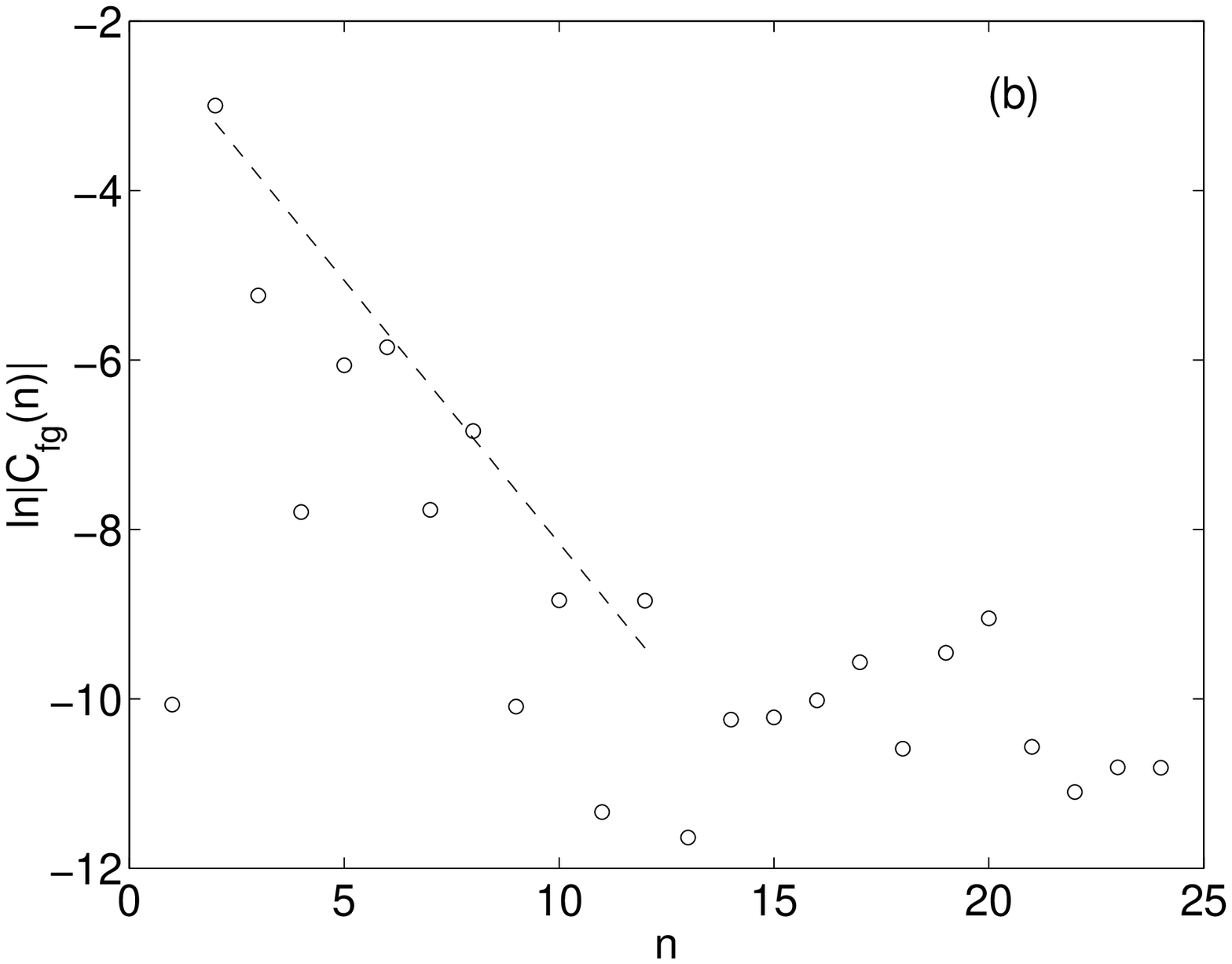} }
\end{minipage}
\\
\begin{minipage}{7.1cm}
\centerline{\epsfxsize 7.0cm\epsfbox{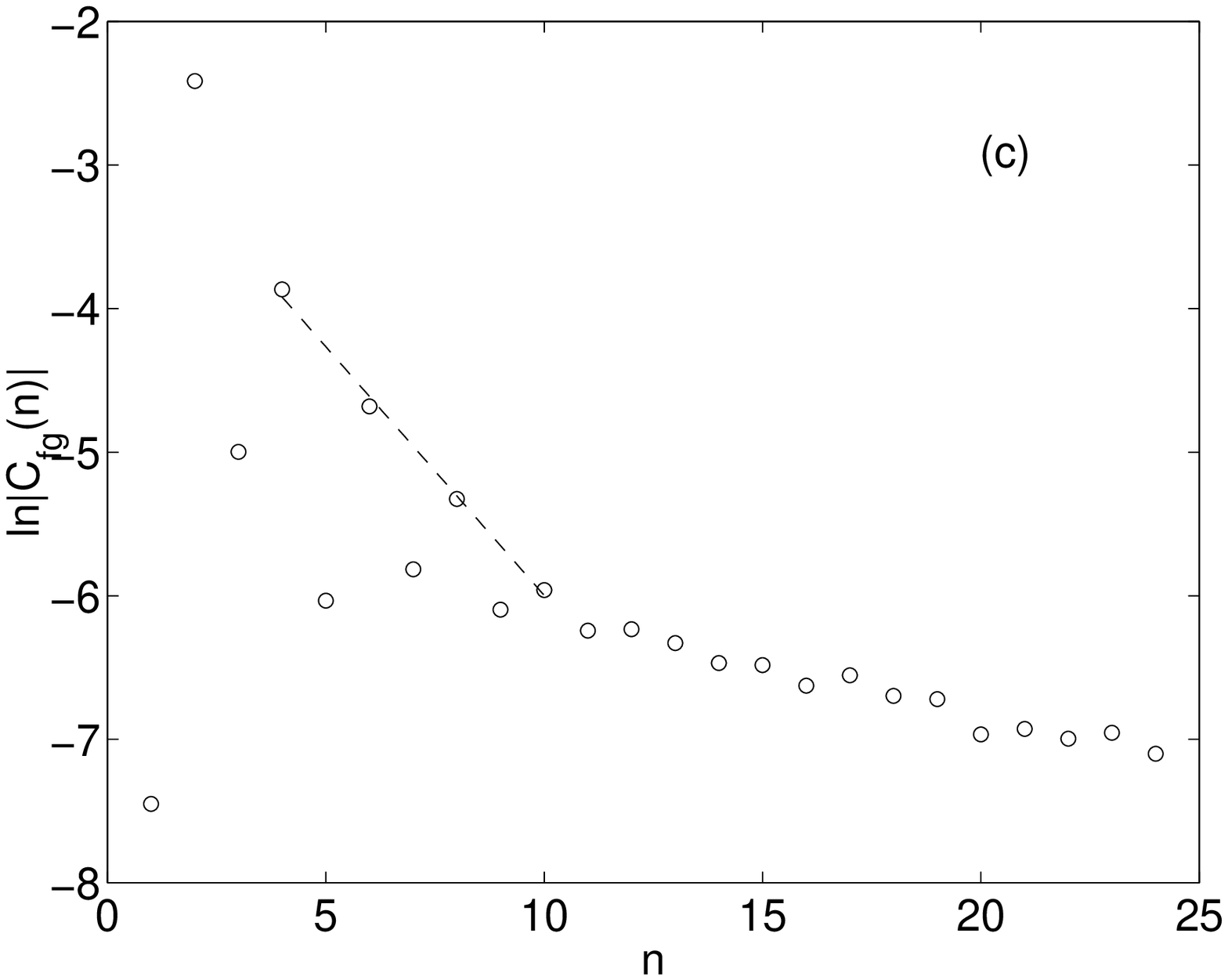} }
\end{minipage}
\hspace{1.0cm}
\begin{minipage}{7.1cm}
\centerline{\epsfxsize 7.0cm\epsfbox{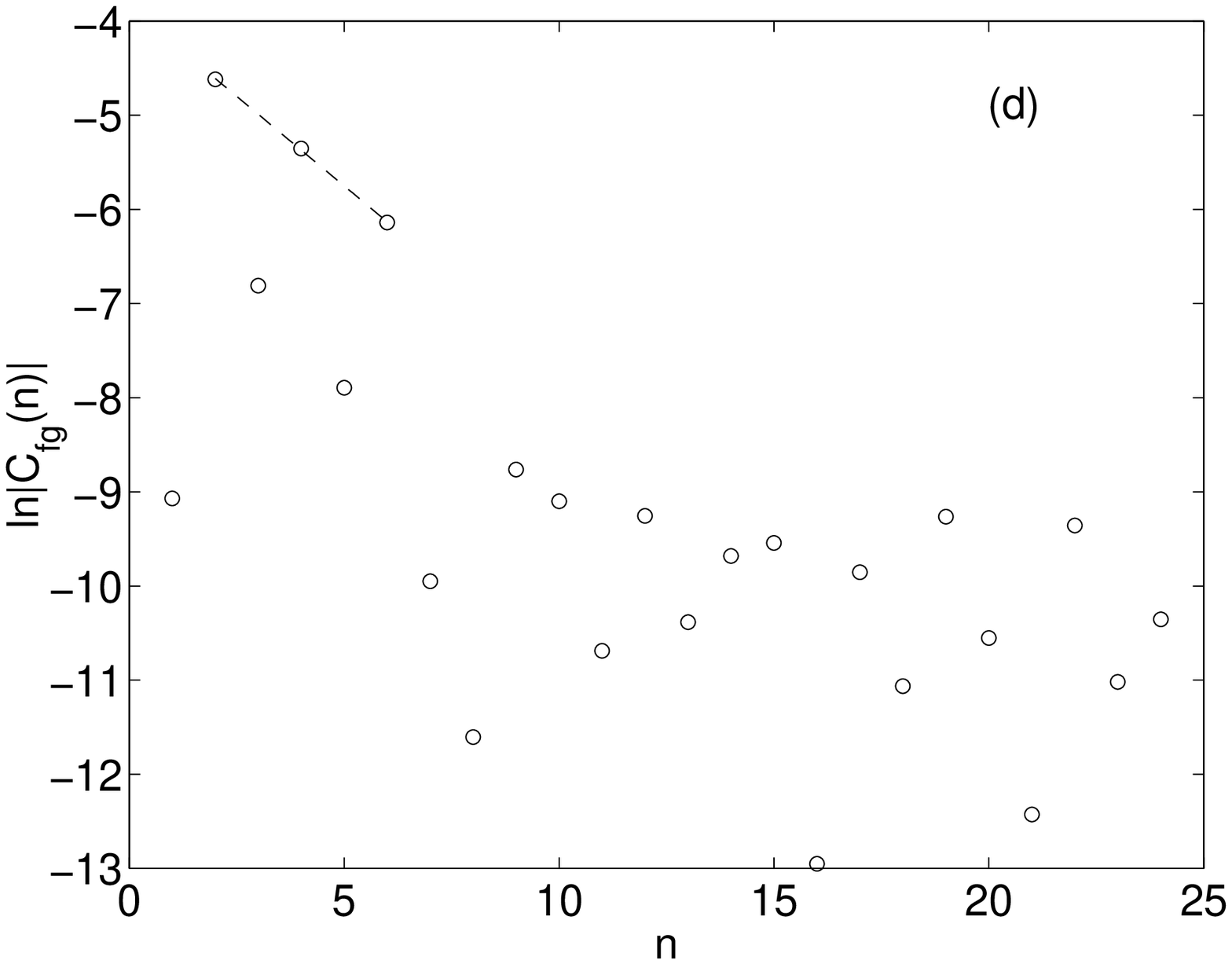} }
\end{minipage}
\hspace{1.0cm}
\end{center}
\caption{The absolute value of the function $C_{fg}(n)$ for 
$f=g=\phi_{01} $ and
(a) K=10.7,
(b) K=14.3, 
(c) K=12.55, 
(d) K=14.7.
The dashed line represents the best fit to the data.
The values $s=1$ and $N=10^{8}$ were used.}
\end{figure}
\begin{figure}
\centerline{\epsfysize 12.0cm \epsfbox{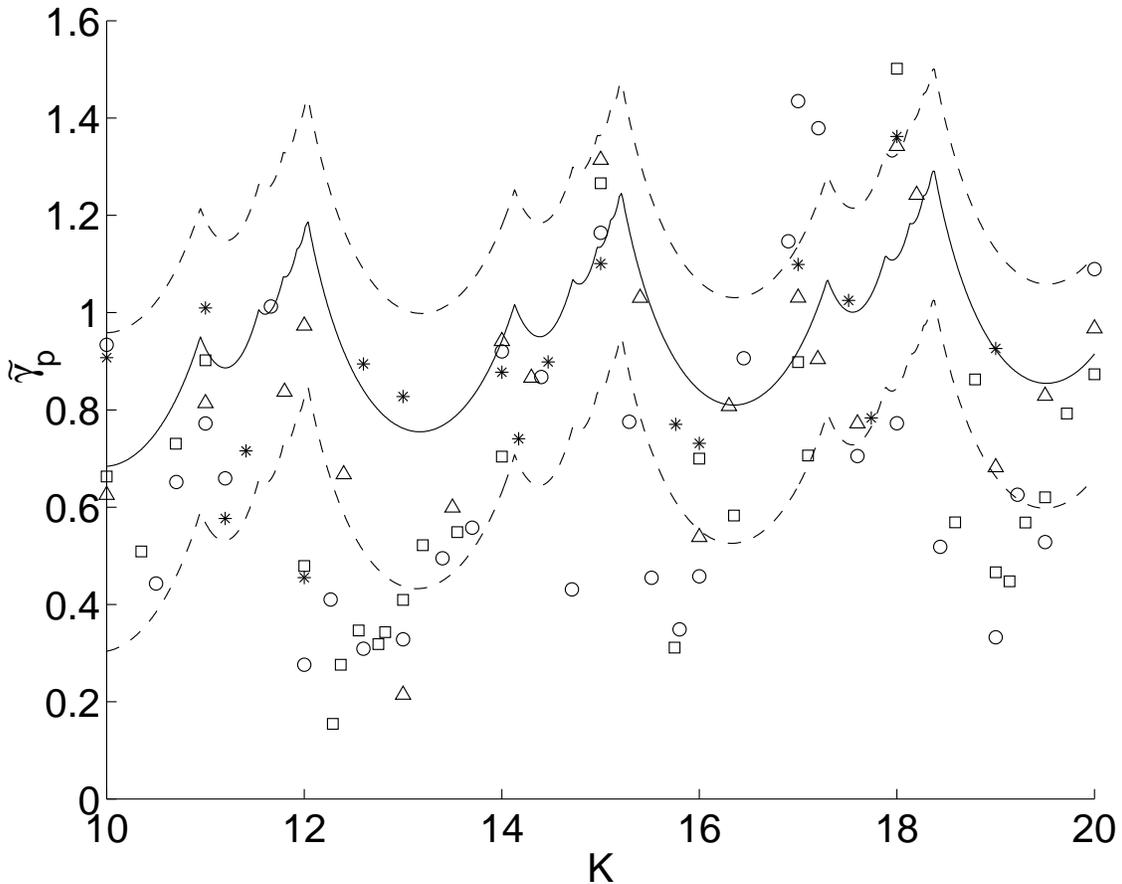}}
\caption{The fast relaxation rate $\tilde{\gamma}$ as found from plots
like Figs. 5 and 6 for $f=\phi_{01}$, $g=\phi_{02}$(triangles),
$f=\phi_{02}$, $g=\phi_{04}$(circles), $f=g=\phi_{02}$(stars) and 
$f=g=\phi_{01}$(squares), compared to the theoretical value 
(\protect\ref{gammap2}) (solid line).
The dashed lines denote the theoretically estimated error.
Here we used $s=1$ and $N=10^{8}$.}
\end{figure}

\section{Summary and Discussion}  
  
Relaxation to equilibrium was studied for the kicked rotor that is a  
standard system for the exploration of classical chaos in driven   
systems and its quantum mechanical suppression. Relaxation and  
diffusion are important concepts in statistical mechanics.  
Here they were studied for a mixed chaotic system. Very 
little is known rigorously about   
such systems although most models describing real   
physical systems are mixed, namely in some regions of phase space   
the motion is   
regular while in some regions it is chaotic.  
  
In this work the kicked rotor was studied in a phase space that is   
the torus defined by (\ref{d3}). The relaxation of distributions in phase   
space takes place in stages. First, the inhomogeneity in $\theta$   
decays with  rapid relaxation rates, the slowest of them  
is $\tilde{\gamma}$. Then   
relaxation of the inhomogeneities in the $J$ direction takes place   
with the relaxation rates related to the diffusion coefficient   
via (\ref{gammak2}). Diffusion   
was previously believed to be a good approximation for the kicked rotor, 
but here, to our best  knowledge, the various time scales 
were analyzed carefully for the first time. In particular we have found the
time scale, $1/\tilde{\gamma}$, below which 
the diffusion approximation does not hold since relaxation of 
correlations in the  angle direction still takes place. 
  
There is a clear relation between the relaxation of inhomogeneities   
in $ \theta $ and the diffusion constant since  
\begin{equation}  
<(J_{n+1}-J_{0})^{2}>=\sum_{i,j=0}^{n} K^{2}<sin\theta_{i}   
sin\theta_{j}>,  
\label{disc1}  
\end{equation}  
where $ J_{i}$ and $ \theta_{i} $ are the momentum and angle before the $i$-th kick.  
For a chaotic trajectory  
\begin{equation}  
<sin\theta_{i} sin\theta_{j}>=<sin\theta_{0} sin\theta_{|i-j|}>=  
C_{ff}(|i-j|),  
\label{disc2}  
\end{equation}  
where $ C_{ff}(|i-j|) $ is the correlation function   
(\ref{cf1}) with  
$ f= \phi_{01} $. If the sum $ \sum_{i=0}^{\infty} C_{ff}(i) $  
converges, as is the case where $C_{ff}$ falls off exponentially,  
diffusion is found  
and the value of the diffusion coefficient is   
\begin{equation}  
D=\frac{K^{2}}{2}\sum_{i=-\infty}^{\infty}C_{ff}(i).  
\label{DRW1}  
\end{equation}  
In App. C we show that (\ref{u25}) that was obtained by   
Rechester and White in ~\cite{RW}  
is just   
\begin{equation}  
D=\frac{K^{2}}{2}\sum_{i=-2}^{2}C_{ff}(i).  
\label{DRW}  
\end{equation}  
A derivation that is very similar is presented in ~\cite{Cary}.  
If the sum diverges one obtains anomalous diffusion.   
   
Finite noise leads to   
the effective truncation of the evolution operator (\ref{start}) . In the basis (\ref{3}) it means that it results in limited resolution. Moreover for $\sigma > 0$ the operator  
$ \tilde{U} $ is nonunitary.   
The approximate eigenvalues of $ \hat{U} $ given by (\ref{start}), that were found  
in this work are $1$ and $z_{k}$ of (\ref{u23})   
(~if (\ref{ad2}) satisfied ) and   
$ \tilde{z}$ of (\ref{gammap1}). In our approximation method   
we could not obtain many eigenvalues related to angular 
relaxation modes. Because of the  
effective truncation,   
$ \psi_{\gamma} $, the eigenfunction of $ \hat{U} $, can be expanded in terms of the basis states (\ref{3}). The relaxation rates of these states are  
$-ln(z_{k})$ and $ -ln(|\tilde{z}|)$, where $ z_{k} $ and   
$\tilde{z}$ are given by (\ref{u23}) and (\ref{gammap1}).  
In the limit $ \sigma \rightarrow 0 $ the evolution operator is unitary,  
$ \psi_{\gamma} $ approach some generalized functions while     
$ z_{k} $ and $\tilde{z}$ approach the values of the poles of the  
matrix elements of the resolvent $ \hat{R} $ of (\ref{u1})  
obtained from the extrapolation from $ |z|>1 $  
(corresponding to the $ 0<\epsilon  
\rightarrow 0 $, used in the standard definition of the Green's function~). 
  
These are the Ruelle resonances that are related to the relaxation rates   
via (\ref{gammak1}) and (\ref{gammap1}). This is very similar to the situation for hyperbolic   
systems such as the baker map. For hyperbolic systems  
the Ruelle resonances ( related to the relaxation rates ) approach  
fixed values inside the unit circle in the complex $z$ plane in the   
limit of an infinite matrix   
for the evolution operator or of infinitely fine phase space   
resolution. This was found to be correct also here when one takes the   
limit   
$ \sigma \rightarrow 0 $ in (\ref{u24}) and (\ref{ex7}) resulting in   
(\ref{resk}) and (\ref{resp}).  
Numerical tests in absence of noise confirm that the analytical results   
provide a good approximation for the relaxation to equilibrium  
and diffusion in the chaotic component.  
Results of similar nature were found in the standard map with  
$s=1$ for some values of $K$  \cite{blum}, for the ``perturbed cat''  
map \cite{blum}, and also for the kicked top \cite{HW}.  
In all these works it was found, within the  
approximations used,  that the leading resonances are either  
real or form the quartet   
$(\pm A, \pm i A)$ where $A$ is a real number satisfying $0<A<1$.  
The generality of this  
form should be subject to further research.  
For the kicked top it was attributed \cite{HW} to the  
dominance of an orbit of period $4$.   
  
In mixed systems, such as the kicked rotor, even in the chaotic components   
there is sticking to regular islands and acceleration modes.  
Noise eliminates this sticking . The analytic formulas (\ref{resk}) and  
(\ref{resp}) are obtained from an expansion in powers of $ 1/ \sqrt{K}$   
for finite variance of noise $\sigma^{2}$ and the limit   
$ \sigma^{2} \rightarrow 0 $ is taken in the end of the calculation.  
A nonvanishing value of $ \sigma^{2} $ assures the convergence of the   
series (\ref{w5}).   Appearance of the islands and the sticking is a  
nonperturbative effect and therefore it is not reproduced in our theory. For   
this reason in absence of noise the results are only approximate.  
The effect of the sticking is extremely small for most values of the   
stochasticity parameter $K$, as verified by the numerical calculations   
without noise.   
  
The physical reason for the decay of correlations  
is, that in a chaotic system, because of the stretching and   
folding mechanisms there is persistent flow in the direction of   
functions with finer details namely larger $|k|$ and $|m|$ in our case.  
Consequently the projection on a given function, for example one of the   
basis functions (\ref{3}) in our case, decays ~\cite{HW,Haake}.   
The crucial point is that this   
function should be sufficiently smooth.  
This argument should hold also for the chaotic   
component of mixed systems. In the present paper the actual relaxation   
rates were calculated.  
Here noise was used in order to make the analytical calculations  
possible. In real experiments some level of noise is present,  
therefore the results in presence of noise are of experimental  
relevance.  
It was shown   
with the help of the Cauchy-Hadamard theorem ~\cite{CA}  
( see discussion following (\ref{u9}) ) that for $ s>>K>>1 $  
exponential relaxation to the invariant density takes place with the   
rate $ \gamma_{1}=D(K)/s^{2} $, where $ D(K) $ is given by (\ref{d1}).  
It was deduced from the radius of convergence for the series of   
the matrix element of $\hat{R'}$ (see (\ref{u6})).  
This rate is independent of $ \sigma $. It is found for all functions   
that can be expanded in the basis (\ref{3}), with an absolutely convergent   
expansion. It excludes for example functions of the form (\ref{s13}).   
We believe this   
statement can be made rigorous by experts.  
  
For the baker map it was found that the resolvent of   
the evolution operator of the   
Quantum Wigner function, when coarse grained has the same poles as the   
classical Frobenius-Perron operator ~\cite{Shmuelinbook}. 
We believe it should hold also   
here. The fact that the Ruelle resonances of the modes of slow relaxation   
are $ z_{k} $ that are identical to the ones of the diffusion operator   
gives additional support to approximations made for the calculation
of the ensemble averaged localization length in ~\cite{Atland}.  
  
Finally, the Ruelle resonances, that were introduced 
and established rigorously for 
hyperbolic systems can be used to describe    
relaxation and transport in the chaotic component of mixed systems.   
Here it was demonstrated for the kicked-rotor.   
  
  
\section{Acknowledgments}  
We have benefited from discussions with E. Berg,   
R. Dorfman, I. Guarneri,   
F. Haake,   
E. Ott, R. Prange, S. Rahav,  J. Weber and M. Zirenbauer.   
We thank in particular   
D. Alonso for extremely  
illuminating remarks and helpful suggestions. This research was supported in part by the  
US--NSF grant NSF DMR 962 4559, the  
U.S.--Israel Binational Science Foundation (BSF), by the Minerva  
Center for Non-linear Physics of Complex Systems, by the Israel  
Science Foundation, by the Niedersachsen Ministry of Science  
(Germany) and by the Fund for Promotion of Research at the  
Technion. One of us (SF) would like to thank R.E. Prange for the  
hospitality at the University of Maryland where this work was  
completed.

\begin{appendix}  
\section{The matrix elements of the evolution operator in presence of noise.}  
\label{appA}  
In this appendix the matrix elements in the representation (\ref{3})  
are calculated.  
For this purpose (\ref{s9}) is transformed to the Fourier representation by  
\begin{equation}  
(k_{2}m_{2}|\hat{U}|k_{1}m_{1})=  
\int_{0}^{2\pi}d\theta\int_{0}^{2\pi s}dJ\int_{0}^{2\pi}  
d\theta'\int_{0}^{2\pi s}dJ'  
\label{s10}  
(k_{2}m_{2}|J\theta)(J\theta|\hat{U}_{noise}|J'\theta')  
(J'\theta'|\hat{U}_{K}|k_{1}m_{1}).   
\end{equation}  
Substitution of (\ref{s7}) and  (\ref{s8}) yields  
\begin{equation}  
(k_{2}m_{2}|\hat{U}|k_{1}m_{1})= \int_{0}^{2\pi}d\theta\int_{0}^{2\pi s}dJ  
\int_{0}^{2\pi}d\theta'  
\label{s11}  
\frac{1}{\sqrt{2\pi}}\frac{1}{\sqrt{2\pi s}}\exp (-im_{2}\theta)  
\exp \left(i\frac{-k_{2}J}{s} \right)  
\end{equation}  
\[\sum_{m}^{} \frac{1}{2\pi}\exp(im(\theta'-\theta))\exp\left(-imJ-\frac{\sigma^{2}}{2}m^{2} \right)  
\frac{1}{\sqrt{2\pi}}\frac{1}{\sqrt{2\pi s}}\exp (im_{1}\theta')  
\exp \left(i\frac{k_{1}}{s}(J+K sin \theta') \right).  
\]  
Terms containing $\theta$ are  $\exp(im(-\theta))\exp (-im_{2}\theta)$.  
Integration over $\theta$ yields $\delta_{m,-m_{2}}$ leading to  
\begin{equation}  
(k_{2}m_{2}|\hat{U}|k_{1}m_{1})=   
\int_{0}^{2\pi s}dJ\int_{0}^{2\pi}d\theta'\frac{1}{\sqrt{2\pi s}}\exp \left(i\frac{-k_{2}J}{s} \right)  
\label{s12}  
\end{equation}  
\[  
\frac{1}{2\pi}\exp(i(-m_{2}\theta')\exp\left(im_{2}J-\frac{\sigma^{2}}{2}m_{2}^{2} \right)  
\frac{1}{\sqrt{2\pi s}}\exp (im_{1}\theta')\exp \left(i\frac{k_{1}}{s}(J+K sin \theta') \right).  
\]  
Integration over $J$ results in $\delta_{k_{2}-k_{1},m_{2}s}$, yielding  
  
\begin{equation}  
(k_{2}m_{2}|\hat{U}|k_{1}m_{1})=\frac{1}{2\pi}\int_{0}^{2\pi}d\theta'  
\exp (-i(m_{2}-m_{1})\theta')  
\exp\left(-\frac{\sigma^{2}}{2}m_{2}^{2} \right)  
\exp \left(i\frac{k_{1}}{s}K sin \theta' \right)\delta_{k_{2}-k_{1},m_{2}s}   
\end{equation}  
Finally with the help of the integral representation for Bessel functions:  
\[   
J_{m}(z)=\frac{1}{2\pi}\int_{0}^{2\pi}d\theta\exp (-im\theta)  
\exp (izsin \theta)   
\]  
one obtains (\ref{start}).  
 
\section{The end terms in strings of the fast modes}  
\label{AppB}

In this appendix,  possible examples for contributions to the end terms
in (\ref{ex4}) are 
presented. 
The left end term is a sum of terms of the form   
\begin{eqnarray}  
& &C^{(l)}(m,m^*)= \nonumber\\  
& &J_{2m+m^*}(-mK)~J_{-m-2m^*}(m^*K)~J_0(0)  
~\exp\left[-\frac{\sigma^2}{2}\left(m^2+(m+m^*)^2+m^{* 2}\right)\right] \nonumber \\   
&+&\sum_{m_1}J_{m-m_1}(-mK)~J_{2m_1+m+m^*}((-m-m_1)K)~J_{-m-2m^*-m_1}(m^*K)~J_0(0)~\nonumber\\  
& &\exp\left[-\frac{\sigma^2}{2}\left(m^2+m_1^2+(m+m^*+m_1)^2+m^{* 2}\right)\right]...  
\label{ex5}  
\end{eqnarray}  
where in the first term $m_1=(m+m^*)$ and $m_2=m_3=m^*$ while in the second term  
$m_2=-(m+m^*+m_1)$ and $m_3=m_4=m^*$. For $q \neq 0$ the right end term is a sum of terms of the  
form  
\begin{eqnarray}  
& &C^{(r)}(m^*,m',q)= \nonumber\\  
& &\sum_{m_{n-1}}J_{2m^*+m_{n-1}+q}(-m^*K)~J_{-m^*-2m_{n-1}-q}((m_{n-1}+q)K)~J_{m_{n-1}-m'}(qK)  
\nonumber \\  
& &~\exp\left[-\frac{\sigma^2}{2}\left(m^{* 2}+(m^*+m_{n-1}+q)^2++m_{n-1}^2\right)\right]....  
\label{ex6}  
\end{eqnarray}  
where $m_{n-2}=-(m^*+m_{n-1}+q)$ and $m_{n-3}=m^*$.  
For $q=0$ one has to take $m_{n-1}=m'$ and the end term consists of a sum over $m_{n-2}$.

\section{The relation between the diffusion coefficient and   
correlation function.}  
\label{AppC}  
  
In this appendix the relation between (\ref{DRW})  
and (\ref{u25}) will be derived  
( for a somewhat similar derivation see ~\cite{Cary} ).  
For this purpose we note that  
\begin{equation}  
C_{ff}(n)=\int_{0}^{2\pi}  
\frac{d\theta}{2\pi} \int_{0}^{2\pi s}\frac{dJ}{2\pi s}  
sin\theta\ \hat{U^{n}}\ sin\theta=  
-\frac{1}{4}[(0,-1|-(0,1|]\hat{U^{n}}[|0,1)-|0,-1)],  
\label{C1}  
\end{equation}  
where the representation $|k,m)$ (see (\ref{3})) is used.  
The matrix elements of $\hat{U}$ are given by (\ref{start})  
and the matrix elements of $\hat{U^{2}}$ required for the present   
calculation are  
\begin{equation}  
(0,m_{2}|\hat{U^{2}}|0,m_{1})=  
J_{2m_{2}}\left(-m_{2}K\right)e^{-\sigma^{2}m_{2}^{2}}  
\delta_{m_{1},-m_{2}}  
\label{C2}  
\end{equation}  
as can be easily obtained from multiplication of two matrices of the   
form (\ref{start}).  
From (\ref{C1}) it is clear that $C_{ff}(0)=\frac{1}{2}$. Inspecting  
(\ref{start}) with $k_{1}=k_{2}=0$ one notes that  it is required  that also  
$m_{1}=m_{2}=0$, therefore $C_{ff}(1)=0$. Substitution of (\ref{C2})  
in (\ref{C1}) yields  
\begin{equation}  
C_{ff}(2)=-\frac{1}{2}J_{2}\left(K\right)e^{-\sigma^{2}}.  
\end{equation}  
Using the fact that $ C_{ff}(-n)=C_{ff}(n) $, substitution of the values of   
$C_{ff}(0)$ and $C_{ff}(2)$ into (\ref{DRW}) yields the   
expression (\ref{u25})  
that was obtained by Rechester and White ~\cite{RW}. Because of the discussion   
following (\ref{w6}) the correlation functions $C_{ff}(n)$ with $n>2$ lead to terms   
that are  of higher orders in $1/\sqrt{K}$ than (\ref{u25}).  
  
\end{appendix}

\typeout{References}

\end{document}